\documentclass[lettersize,journal]{IEEEtran}
\usepackage{amsmath,amsfonts}
\usepackage{algorithmic}
\usepackage{algorithm}
\usepackage{array}
\usepackage{textcomp}
\usepackage{stfloats}
\usepackage{url}
\usepackage{verbatim}
\usepackage{graphicx}
\usepackage{cite}

\usepackage{lineno,hyperref,multirow,diagbox,multicol,balance,epstopdf, subfigure}
\usepackage[numbers]{natbib}
\usepackage{booktabs}
\usepackage{bm}
\usepackage{threeparttable}
\usepackage{color, xcolor,makecell, array}

\begin{document}

\title{ParaTTS: Learning Linguistic and Prosodic Cross-sentence Information in Paragraph-based TTS}

\author{Liumeng Xue,
Frank K. Soong,
Shaofei Zhang, 
Lei Xie$^{*}$
\thanks{Corresponding author: Lei Xie}
\thanks{Liumeng Xue and Lei Xie are with the Audio, Speech and Langauge Processing Group (ASLP@NWPU), School of Computer Science, Northwestern Polytechnical University, Xi’an 710072, China. Email: lmxue@nwpu-aslp.org, lxie@nwpu.edu.cn}
\thanks{Frank K. Soong is with Microsoft Research Asia (MSRA), Beijing 100080, China. Email: frankkps@microsoft.com}
\thanks{Shaofei Zhang is with Microsoft Azure Speech, Beijing 100080, China. Email: shazh@microsoft.com). }
\thanks{This work was supported by the National Key Research and Development Program of China (No. 2020AAA0108600).}
}



\maketitle

\begin{abstract}
Recent advancements in neural end-to-end TTS models have shown high-quality, natural synthesized speech in a conventional sentence-based TTS. However, it is still challenging to reproduce similar high quality when a whole paragraph is considered in TTS, where a large amount of contextual information needs to be considered in building a paragraph-based TTS model. To alleviate the difficulty in training, we propose to model linguistic and prosodic information by considering cross-sentence, embedded structure in training. Three sub-modules, including linguistics-aware, prosody-aware and sentence-position networks, are trained together with a modified Tacotron2. Specifically, to learn the information embedded in a paragraph and the relations among the corresponding component sentences, we utilize linguistics-aware and prosody-aware networks. The information in a paragraph is captured by encoders and the inter-sentence information in a paragraph is learned with multi-head attention mechanisms. The relative sentence position in a paragraph is explicitly exploited by a sentence-position network. Trained on a storytelling audio-book corpus (4.08 hours), recorded by a female Mandarin Chinese speaker, the proposed TTS model demonstrates that it can produce rather natural and good-quality speech paragraph-wise. The cross-sentence contextual information, such as break and prosodic variations between consecutive sentences, can be better predicted and rendered than the sentence-based model. Tested on paragraph texts, of which the lengths are similar to, longer than, or much longer than the typical paragraph length of the training data, the TTS speech produced by the new model is consistently preferred over the sentence-based model in subjective tests and confirmed in objective measures.

\end{abstract}


\section{Introduction}
The development of sequence-to-sequence (seq2seq) based neural acoustic models~\cite{ Wang2017TacotronTE,shen2018natural,Li2019NeuralSS, Ren2019FastSpeechFR,Ren2020FastSpeech2F,Yu2020DurIANDI} and neural vocoders ~\cite{Oord2016WaveNetAG,Valin2019LPCNETIN,Kumar2019MelGANGA,Kong2020HiFiGANGA20hifi}, brought a significant improvement to the text-to-speech (TTS) synthesis quality, which allows to automatically synthesize human-like natural speech with high fidelity. TTS is widely applied to many scenarios, such as voice assistant, navigation, smart customer service, audiobook, just to name a few, due to its high efficiency and low cost compared to manual recordings. The audiobook is an important TTS application in rendering the text of stories into expressive voices.

The text of an audiobook is composed of successive sentences, paragraphs, sections and chapters in a coherent and hierarchical form to describe a fictional story or thoughts of an author~\cite{granville1993algorithm, hearst1994multi}. Moreover, such a coherent and hierarchical relationship in the text has an impact on how it is being uttered in human voice formation. A paragraph, consisting of one or more sentences is a self-contained unit of a discourse in writing for conveying a particular point or idea. Prosodic patterns in paragraph audios have been observed~\cite{ Smith2004TopicTA,Tseng2006ProsodicFA, Cole2015ProsodyIC}. For instance, pitch resets --- higher pitch and increased pitch range are usually observed at the beginning of the new paragraph. Similar reset pattern has also been found in energy~\cite{Kreiman1982PerceptionOS} or RMS amplitude~\cite{Grosz1992SomeIC,Hirschberg1996APA,Swerts1994ProsodyAA}. Declination --- the tendency of pitch and energy to decline over the paragraph, ending with low pitch and energy. Lengthening ---  speech rate peaks in the middle of a paragraph, lengthening in the initial and final position.

To synthesize speech on a paragraph basis, a straightforward way is to synthesize each sentence in a paragraph and then combined them together. A similar approach has been used in news and navigation broadcast applications~\cite{alias2005high}. However, there are three disadvantages: (i) an additional step of post-processing is necessary to integrate individual sentences; (ii)  during combining sentences, break duration length between two consecutive sentences should be considered to ensure the integrated paragraph speech sounds natural. Pause plays a significant role in the storytelling and pause prediction is studied for early statistical parametric speech synthesis (SPSS)~\cite{sarkar2015modeling, sarkar2015data, sarkar2015analysis}; (iii) the prosody in the combined paragraph speech may become inconsistent and not smooth perceptually, varying greatly from one sentence to the next and resulting in unnatural transitions. This issue can be alleviated by minimizing the acoustic variation and linguistic distance between a sentence and the previous one~\cite{tyagi2020dynamic}. All these issues make the process of paragraph speech synthesis complex and challenging. Furthermore, it is not a suitable method for paragraph-level speech synthesis because prosodic and acoustic difference exists between sentences spoken in isolation or in a paragraph~\cite{zhang2004prominence, prahallad2006sub, bennett2005prediction, miller1998pronunciation}. An alternative solution is to synthesize speech at a paragraph level directly. It is feasible in the current mainstream seq2seq TTS frameworks for long-form speech synthesis~\cite{Battenberg2020LocationRelativeAM, Wang2020sTransformerSF}, but it still tends to bring listening fatigue to the listeners because the model does not provide appropriate prosody information in a paragraph. Evidence has shown that paragraph-level or discourse-level prosody improves the naturalness and expressiveness of SPSS~\cite{Hu2016DiscoursePA, Peir-Lilja2018}, but annotations of discourse structure or prosodic properties are required. Preparing a large, well-annotated labeling speech dataset is usually too time-consuming and expensive to be practical.

In this paper, we aim at building a natural and expressive paragraph-based end-to-end TTS in a data-driven way. Constrained by the memory and computing resources, we train the model sentences-wise, incorporating paragraph-based contextual information into the training process. Without discourse-level structure annotation, we try to learn paragraph linguistic knowledge from the text itself. Specifically, we adopt a paragraph text encoder to extract high-level paragraph linguistic representation. Regarding prosody, we use typical prosody features including pitch, energy and duration extracted from the corresponding paragraph audios and capture paragraph prosodic information via a paragraph prosody encoder. Meanwhile, we apply multi-head attention mechanisms to learn inner linguistic and prosodic relations between sentences with their paragraph, a sentence-position network is used to further enhance the relevant context of sentences in the paragraph.

The contributions of this paper are summarized as follows:
\begin{itemize}
    \item This paper presents, to our knowledge, the first attempt to build a paragraph-based, end-to-end TTS with the linguistic and prosodic knowledge learned in a data-driven way. The sentence-position network adopted in our model provides simple yet effective features of individual sentences in multi-sentence paragraph generation and improves the naturalness of the synthesized paragraph.
    \item Experimental results show that our proposed model can achieve good performance for generating natural, good-quality paragraph-based speech. We demonstrate that the proposed model can also learn more accurate pause duration between consecutive sentences and has a good generalization capability of producing natural speech for a given long paragraph which can be longer, or much longer than the paragraph used in the training data. 
    \item We also find that subjectively it is rather difficult to evaluate a long paragraph due to the relative short memory in a sequential audio listening test, which is indicated early in~\cite{Clark2019EvaluatingLT}.
    
    
\end{itemize}

\section{Related works}
\textbf{Paragraph-related works}
A paragraph is a self-contained unit of a discourse in writing dealing with a particular point or idea, and the paragraph in spoken discourse carries a variety of information. Discourse relations (DR) expressing how different segments (i.e. elementary discourse units) of a text are logically connected have been studied and used to improve the naturalness of statistical parametric speech synthesis (SPSS) at sentence level~\cite{Aubin2019ImprovingSS}. In addition to discourse structure itself, some works have studied the correlations between discourse and prosody~\cite{Farrs2016ParagraphbasedPC}, demonstrating that discourse structure contributes to overall discourse prosody. Furthermore, the discourse structure and prosody were applied to Hidden Markov Model (HMM) based TTS to improve the prosody of passage synthesized speech~\cite{Hu2016DiscoursePA}. Moreover, in addition to the intra-paragraph prosody patterns, inter-paragraph prosody patterns have also been investigated and then implemented into SPSS to improve the naturalness of the synthesized articles' speech~\cite{Peir-Lilja2018}. Our work differs from these in the following two aspects: (i) they used SPSS  while we use the current mainstream end-to-end TTS approach; (ii) they need discourse relation or prosody annotations for TTS modeling while we do not need any annotations for model training.

Recently, a chapter-wise understanding system for TTS in Chinese novels~\cite{pan2021chapter} is related to our work, in which the chapter-wise understanding system realizes two text understanding tasks in Chinese novels - speaker determination and emotion classification for various voices and emotional expressions speech synthesis. The differences between this and ours are that: (i) they mainly focus on speakers and emotions from the text while we focus on linguistic and prosodic knowledge from both text and the corresponding audios; (ii) they capture information at the chapter level for audiobook speech synthesis while we learn information at the paragraph level for paragraph speech synthesis.

\textbf{Context-related works}
There has been a wide range of research focused on learning or extracting contextual information to improve the performance of TTS. Multiple studies used textual context information extracted from text to improve the sentence prosody~\cite{hodari2021camp, karlapati2021prosodic} or cross-sentence prosody~\cite{xu2021improving} for sentence-based speech synthesis, or capture conversation information for conversational speech synthesis~\cite{guo2021conversational}. Specifically, the textual context information can be semantics-related features extracted by pre-trained models~\cite{xu2021improving, guo2021conversational}, i.e., BERT~\cite{kentonbert} or syntax-related features represented by parse trees~\cite{Guo2019,karlapati2021prosodic} or statistics~\cite{guo2021conversational,hodari2021camp}. Apart from the textual context, it has been reported that acoustic features from the previous sentence can also lead to improvement of sentence-based TTS~\cite{oplustil2020using}. Following this result, a comparison investigation on multiple context representation types of the previous sentence was studied~\cite{oplustil2021comparing}, including textual and acoustic features, utterance-level and word-level features, and representations extracted with a large pre-trained model and learned jointly with the TTS training. Our work differs from these in two folds: (i) the contextual information used in these models was derived from either isolation sentences or consecutive sentences with a predefined length, while in our work, the contextual information is extracted from a variable length paragraph which is a self-contained unit of discourse and composed of several connective sentences; (ii) cross-sentence linguistic context were used to improve sentence-level or conversation-level speech synthesis in their models. By contrast, both linguistic and prosodic cross-sentence context information is used in our work to improve paragraph-level speech synthesis.

		\begin{figure*}[h]
		\subfigure[Paragraph-based TTS model]{
			\begin{minipage}[t]{0.3\linewidth}
				\centerline{\includegraphics[scale=0.47]{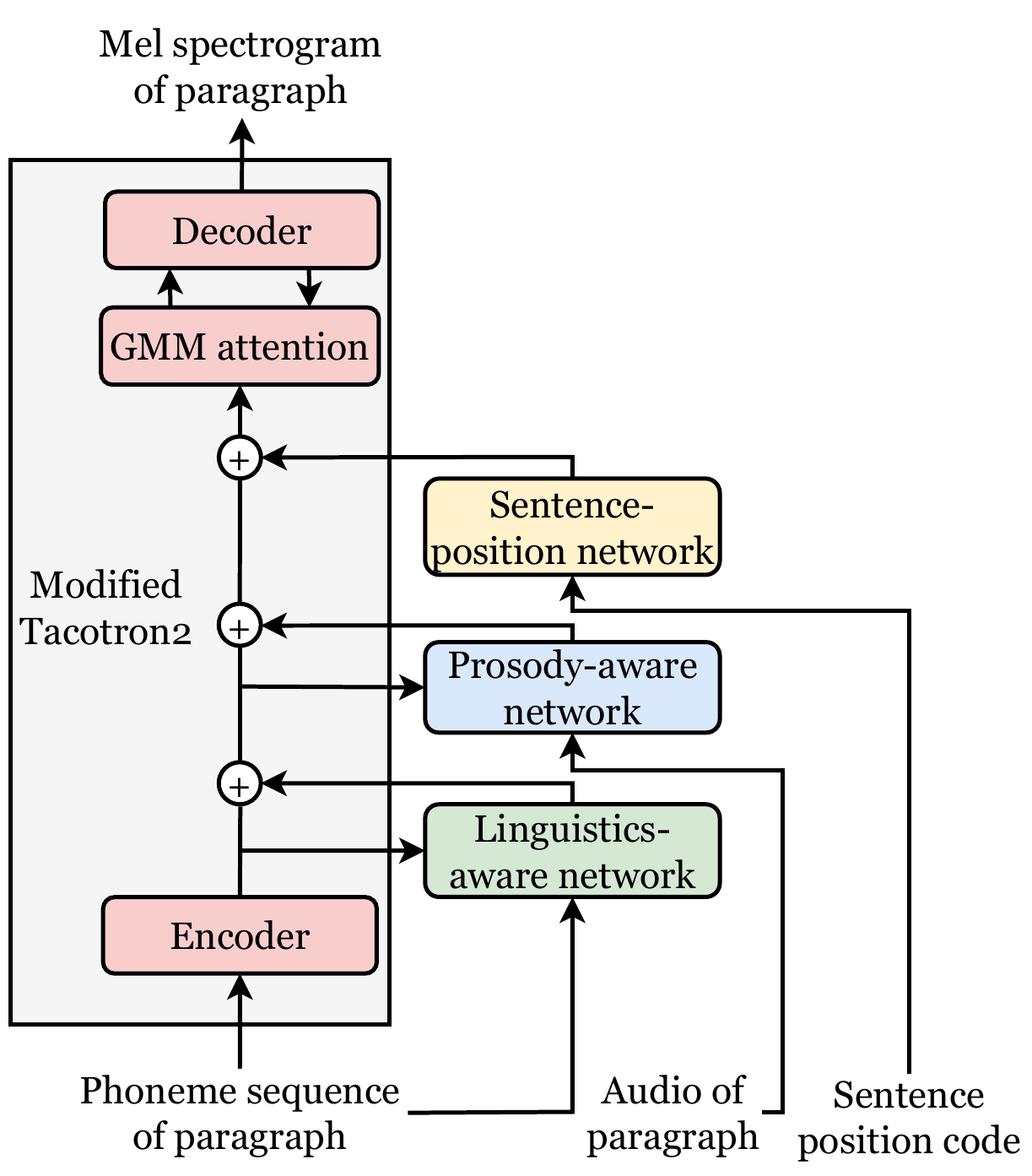}}
		\end{minipage}}
		\hfill
		\subfigure[Linguistics-aware network]{
			\begin{minipage}[t]{0.19\linewidth}
				\centerline{\includegraphics[scale=0.47]{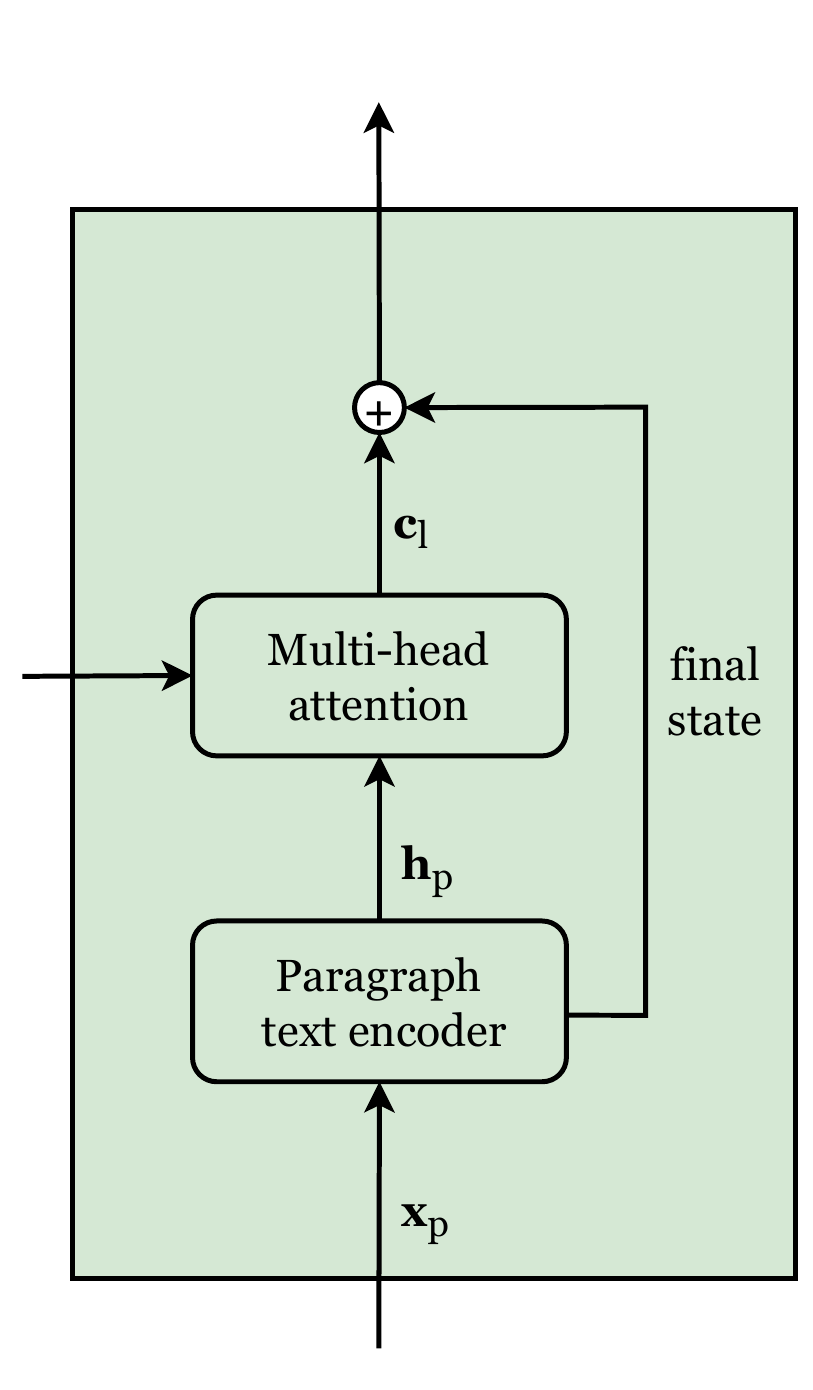}}
		\end{minipage}}
		\hfill
		\subfigure[Prosody-aware network]{
			\begin{minipage}[t]{0.25\linewidth}
				\centerline{\includegraphics[scale=0.47]{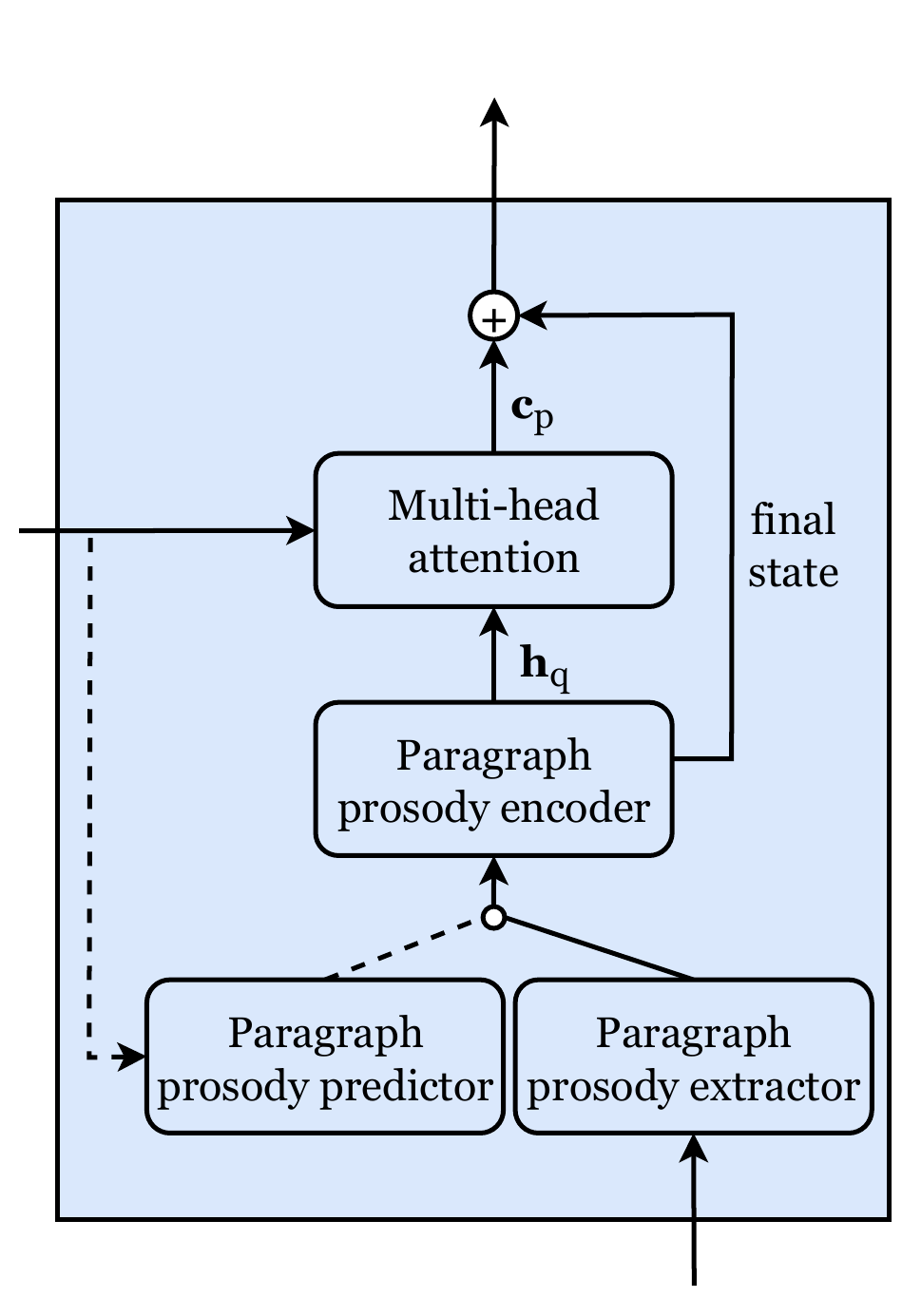}}
		\end{minipage}}
		\hfill
		\subfigure[Sentence-position network]{
			\begin{minipage}[t]{0.19\linewidth}
				\centerline{\includegraphics[scale=0.47]{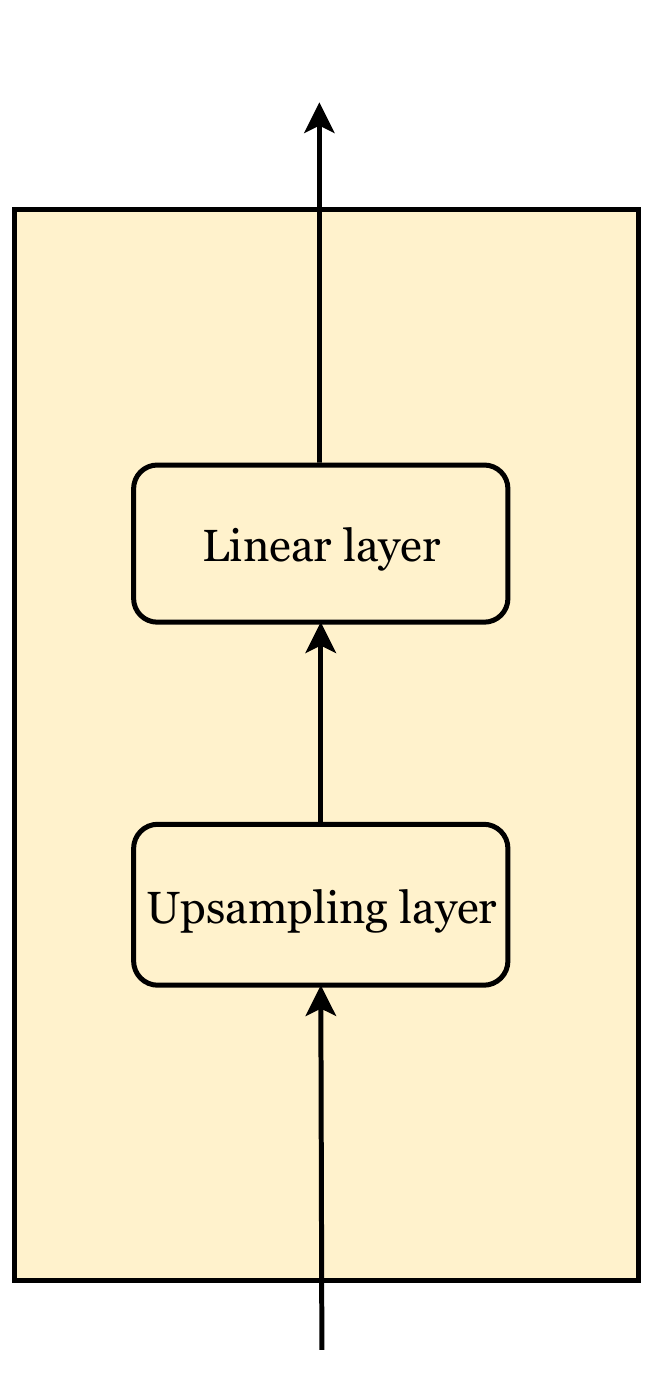}}
		\end{minipage}}
		\hfill
		\caption{The overall architecture for the proposed paragraph-based TTS model (a) Paragraph-based TTS model and three components: (b) Linguistics-aware network. (c) Prosody-aware network. (d) Sentence-position network.}  \vspace*{-5pt}
		\label{fig:paratts}
	\end{figure*}

\section{The proposed model} 
\label{sec:proposed_model}
The architecture of the proposed paragraph TTS model, or ParaTTS in short, is presented in Fig.~\ref{fig:paratts}. It consists of a modified Tacotron2 as the TTS backbone to generate mel spectrograms from a given phoneme sequence, a linguistics-aware network and a prosody-aware network to learn the linguistic and the prosodic knowledge of the whole paragraph and the relationship between sentences and the paragraph, and a sentence-position network to enhance the correlation context of sentences and its paragraph.

In this work, phoneme sequences are used as text inputs \footnote{For Chinese, one Chinese character corresponds to one syllable, and one syllable is composed of at least one phonemes and can correspond to one or more Chinese characters.}. Given a text, i.e., a Chinese character sequence, it is converted to the corresponding phoneme sequence by a front-end module which includes text normalization, part-of-speech tagging and grapheme-to-phoneme conversion. If not stated otherwise, we directly use the phoneme sequences as the inputs of the model for simplicity.

\subsection{The Modified Tacotron2}
\label{subsec:taco2}
The architecture of the modified Tacotron2 is an attention-based encoder-decoder TTS model. Different from the vanilla Tacotron2~\cite{shen2018natural}, we use a CBHG~\cite{lee2017fully} encoder instead of an LSTM encoder because the former is a powerful module for extracting representations from sequences. The encoder is composed of a phoneme embedding layer, a pre-net of 2 fully connected layers and a CBHG. The CBHG is composed of a bank of 1-D convolutional filters, followed by a highway network~\cite{srivastava2015highway} and a bidirectional gated recurrent unit (GRU)~\cite{chung2014empirical} recurrent neural net (RNN). Moreover, we adopt the GMMv2b attention mechanism ~\cite{Battenberg2020LocationRelativeAM}. The GMMv2b is robust for long sequences because it alleviates occasional catastrophic attention failures, such as repeating or skipping. 

The encoder input is the phoneme sequence of a sentence $\textbf{x}_{s}$ in training or the phoneme sequence of a paragraph  $\textbf{x}_{p}$ in inference. Although the different input, the way to be processed is the same. Given a phoneme sequence \textbf {x} = ($x_{1}$, $x_{2}$, $\cdots$ , $x_{n}$), n is the length of the phoneme sequence, the CBHG encoder encodes it into a hidden state \textbf h = ($h_{1}$, $\cdots$ , $h_{n}$) :
\begin{equation}
   \textbf{h} = encoder(\textbf{x})
\end{equation}
where \textbf{h} represents the extracted high-level phoneme representation. Then the decoder generates the current output $s_{t}$ conditioned on the previous prediction at each step:
\begin{equation}
   s_{t} = decoder(s_{t-1}, y_{t-1}, c_{t})
\end{equation}
where $c_{t}$ is the context vector calculated by the attention mechanism which encourage the decoder to attend into the important encoder hidden states when generating the output: 
\begin{equation}
   c_{t} = attention(s_{t-1},\textbf{h})
\end{equation}
Thus the speech sequence \textbf{y} = ($y_{1}$, $\cdots$ , $y_{T}$)  is generated from the input phoneme sequence \textbf{x} based on conditional probability p ($y_{1}$, $\cdots$ , $y_{T}$ $|$ $x_{1}$, $\cdots$ , $x_{n}$), and the conditional probability can be formulated as:
\begin{equation}
   p(\textbf{y} | \textbf{x} ) = \prod_{t=1}^{T} p(y_{t}|{y_{1}, \cdots , y_{t-1}}, \textbf{x})
\end{equation}
\begin{equation}
  p(y_{t}|{y_{1}, \cdots , y_{t-1}}, \textbf{x}) = f(s_{t})
\end{equation}
A linear projection function f is used to predict acoustic features directly based on decoder outputs \textbf{s}.

\begin{figure*}[t]
    \begin{center}
     \includegraphics[scale=0.9]{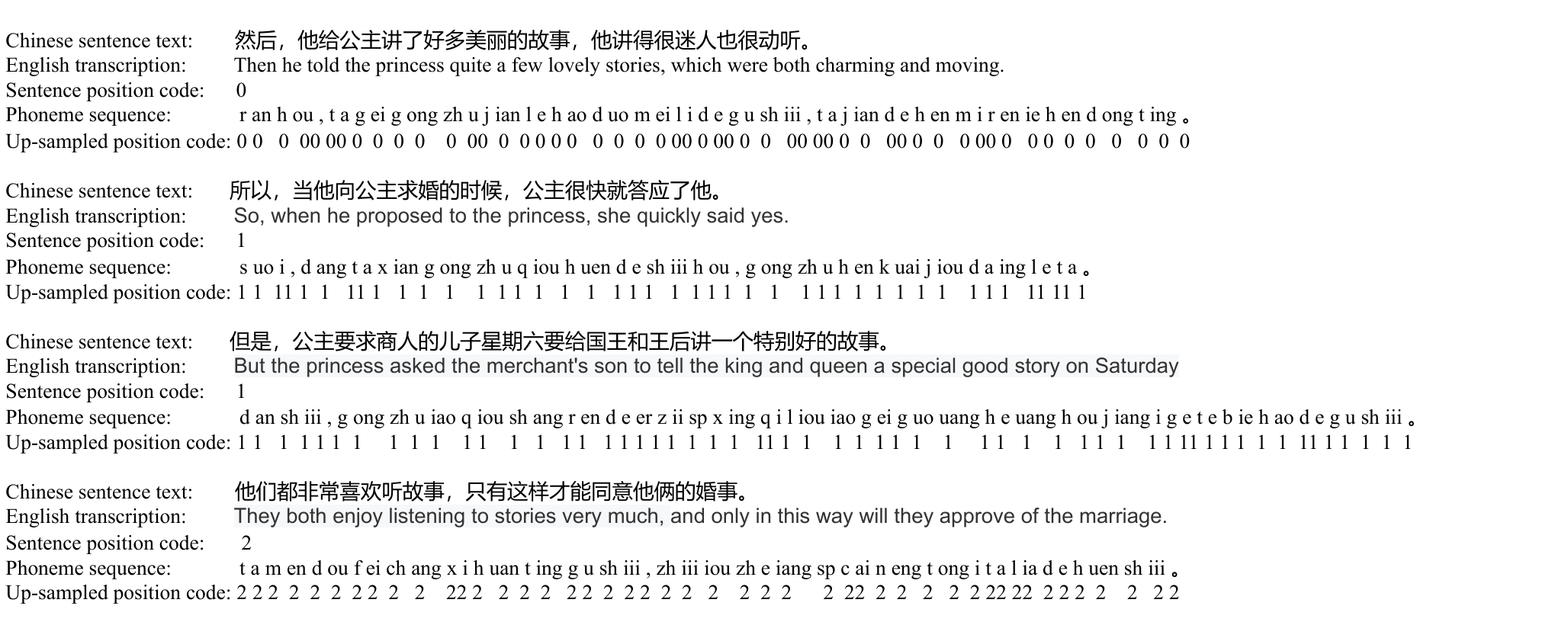}
    \end{center}
    \caption{An example of the sentence position code. There are 4 sentences in the paragraph. The position codes of the first sentence and the last sentence are encoded with codes 0 and 2, respectively, and the middle sentences in between are encoded with code 1.}  \vspace*{-5pt}
    \label{fig:sent_pos}	
\end{figure*}	

\subsection{The Linguistics-aware Network}
The linguistics-aware network consists of a paragraph text encoder with a multi-head attention mechanism, designed to learn the linguistic information of the entire paragraph and the relationship between the component sentence in a paragraph. The paragraph text encoder is also a CBHG as described in \ref{subsec:taco2}, which is used to extract high-level paragraph linguistic representation, $\textbf h_{p} $= ($h_{p1}$, $h_{p2}$, ..., $h_{pm}$), from the given paragraph phoneme sequence, $\textbf{x}_{p}$ = ($x_{p1}$, $x_{p2}$, $\cdots$ , $x_{pm}$), m is the length of the paragraph phoneme sequence. 

In the attention mechanism, attention scores reflect the importance of the key vector with respect to the query vector, allowing the query vector to concentrate on the parts of the key vector~\cite{bahdanau2014neural}. Inspired by this, a attention is employed in this work to capture the important parts of the paragraph phoneme sequence for each sentence phoneme sequence, providing an intrinsic characterisation between a sentence and its paragraph, and meanwhile compensating the lack of paragraph-related information when training model at each sentence unit. Furthermore, the multi-head attention mechanism~\cite{Vaswani2017AttentionIA} is utilized to explore the dependencies in different representation subspaces of the vectors. During training, the sentence hidden representation, $\textbf h_{s} $= ($h_{s1}$, $h_{s2}$, ..., $h_{sn}$), which is encoded from the sentence phoneme sequence $\textbf{x}_{s}$ = ($x_{s1}$, $x_{s2}$, $\cdots$ , $x_{sn}$) (n is the length of the sentence phoneme sequence) via the text encoder in the modified Tacotron2, is used as the query vector \textbf {Q}, and the paragraph hidden representation $\textbf h_{p}$ from the encoder of the linguistic-aware network is used as the key vector \textbf {K} and value vector \textbf {V}. Thus, a linguistic context vector $\textbf{c}_{l}$ = ($c_{l1}$, $c_{l2}$, $\cdots$ , $c_{ln}$), representing the linguistic correlation between the sentence and the paragraph, is calculated as follows:
	\begin{equation}
		\label{equ:lcv}
		\begin{split}
			\textbf {c}_{l} &= multihead(\textbf{Q}, \textbf{K},  \textbf{V}) \\ &= concat(head_{1}, ..., head_{h})\textbf{W}_{O}
		\end{split}
	\end{equation}
	
	\begin{equation}
		\label{equ:ha}
		head_{i} = attention(\textbf{Q}\textbf{W}_{i}^{Q}, \textbf{KW}_{i}^{K},  \textbf{VW}_{i}^{V})
	\end{equation}
	
	\begin{equation}
		\label{equ:af}
		attention(\textbf{Q}, \textbf{K},  \textbf{V}) = softmax(\frac{\textbf{Q}\textbf{K}^{T}}{\sqrt{d_{k}}})\textbf{V}
	\end{equation}
where h is the head number of the multi-head attention mechanism, and i is the i-th head. The $\textbf{W}^{O}$, $\textbf{W}^{Q}$, $\textbf{W}^{K}$ and $\textbf{W}^{V}$ are different projection matrices for output, query, key and value, and the $d_{k}$ is the dimension of the vector \textbf {K}. 
	
The final state of the last bidirectional GRU layer in the paragraph text encoder accumulates the information forward and backward through the whole paragraph phoneme sequence. That is $h_{pm}$, the last element of the paragraph linguistic representation $\textbf h_{p} $= ($h_{p1}$, $h_{p2}$, ..., $h_{pm}$). We view it as a condensed paragraph linguistic representation and add it with the context vector $\textbf{c}_{l}$ as the output of the linguistic-aware network. The output contains the linguistic information of the paragraph and the relationship between the sentence and its paragraph. Finally, we add the output and the encoder output in the TTS backbone model to make the model aware of the linguistic knowledge related to paragraphs.
	
\subsection{The Prosody-aware Network}
The prosody-aware network consists of  4 submodules: paragraph prosody extractor, paragraph prosody predictor and paragraph prosody encoder with a multi-head attention mechanism.

The input of the paragraph prosody encoder is a 3-dimensional, phoneme-level, paragraph prosody feature vector. In training, these features are extracted from the corresponding paragraph speech by the paragraph prosody extractor. In inference, prosody features are predicted by the paragraph prosody predictor which is a 64-unit GRU layer and a 3-unit dense layer. The 3-dimensional prosody features include mean-variance normalized logarithmic fundamental frequency (LF0), intensity, and duration. The paragraph prosody encoder is similar to but different from the reference encoder depicted in~\cite{SkerryRyan2018TowardsEP}. It consists of 6 convolution layers with batch normalization, followed by a 128-unit GRU layer. Here, we use the output of each time step in the GRU as a variable-length paragraph prosody representation, denoted as $\textbf{h}_{q}$ = ($h_{q1}$, $h_{q2}$, ..., $h_{qm}$), which has the same length (m) as the paragraph phoneme representation because the prosody feature input is at the phoneme level. 

Similar to the linguistics-aware network, a multi-head attention mechanism is also used to associate the component sentences with the corresponding paragraph. Specifically, the query vector \textbf{Q} is the addition result of the encoder output of the TTS backbone and the output of the linguistics-aware network. 
The key vector \textbf{K} and the value vector \textbf{V} are both the paragraph prosody representation $\textbf{h}_{q}$. Accordingly, the prosodic context vector $\textbf{c}_{p}$ = ($c_{p1}$, $c_{p2}$, $\cdots$ , $c_{pn}$), representing the prosodic relationship between a sentence and its parent paragraph, can be calculated by the Eqs.~\ref{equ:lcv}-\ref{equ:af}. 
	
Additionally, the final state of GRU in the paragraph prosody encoder, i.e., the last element $h_{qm}$ of the paragraph prosody representation $\textbf{h}_{q}$ = ($h_{q1}$, $h_{q2}$, ..., $h_{qm}$), can be viewed as a compressed paragraph prosody representation. We add it with the prosodic context vector $\textbf{c}_{p}$ to form the output of the prosody-aware network. The output represents the prosodic knowledge related to the paragraph and relations among component sentences in the paragraph. To make the TTS model aware of the prosodic information, we add it with the encoder output in the TTS backbone model.

In inference, the trained paragraph prosody predictor is used to predict the 3-dimensional, phone-level prosody features of the paragraph. The predictor input is the addition result of the linguistic-aware network and the encoder output of the TTS model. We conjecture that the input can predict the paragraph-level prosody features since it contains rich paragraph-relevant information. We train the paragraph prosody predictor using L1 loss between predicted and extracted prosody features and stop gradient flow to ensure the prosody prediction error does not affect the linguistics-aware network and the encoder of the TTS backbone.

	\begin{figure*}[t]
		\subfigure[The distribution of the number of sentences in a paragraph.]{
			\begin{minipage}{0.3\linewidth}
				\centerline{\includegraphics[scale=0.33]{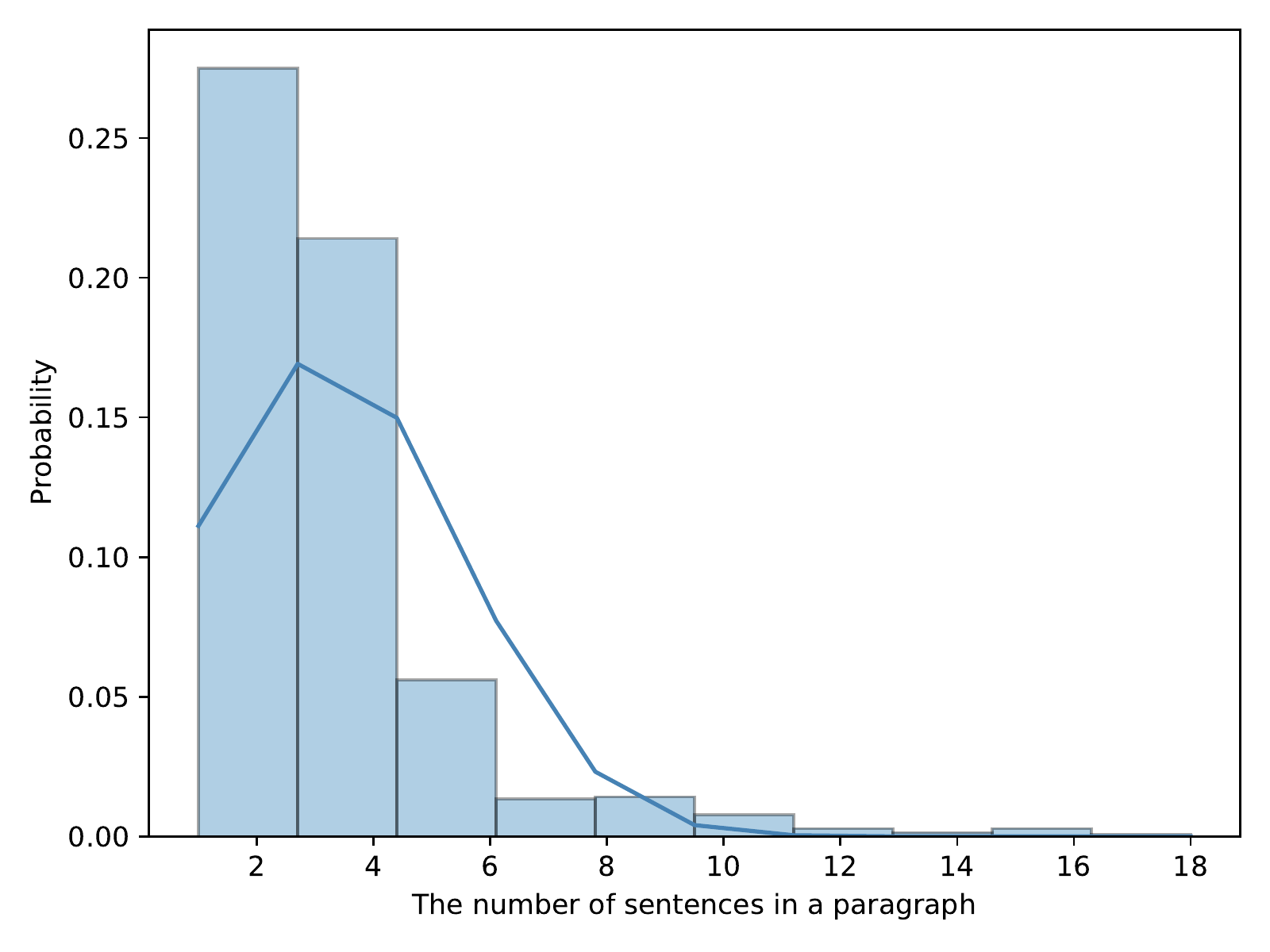}}
		\end{minipage}}
		\hfill
		\subfigure[The distribution of the number of Chinese characters in a sentence. ]{
			\begin{minipage}{0.3\linewidth}
				\centerline{\includegraphics[scale=0.33]{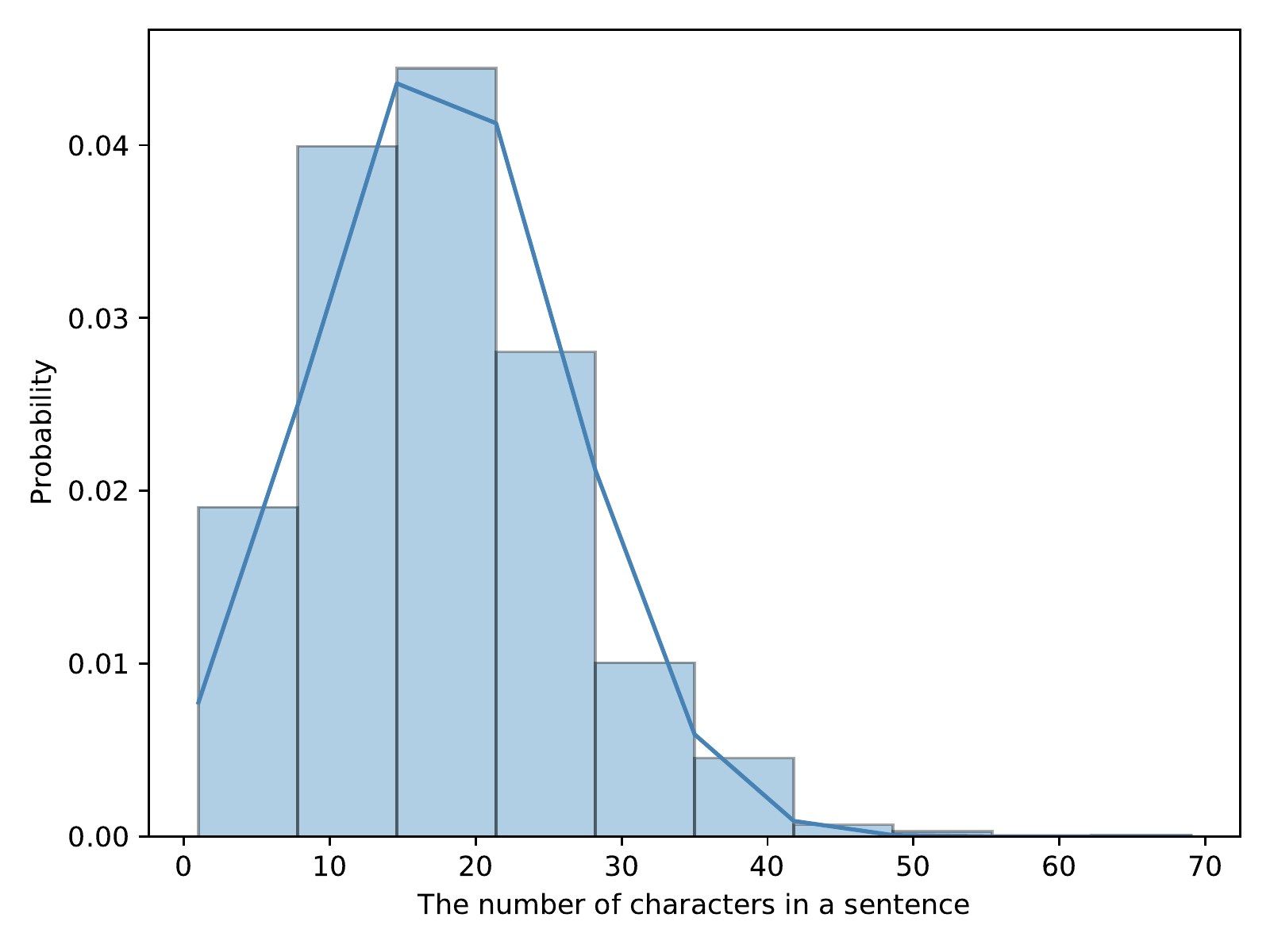}}
		\end{minipage}}
		\hfill
		\subfigure[The distribution of the number of Chinese characters in a paragraph. ]{
			\begin{minipage}{0.3\linewidth}
				\centerline{\includegraphics[scale=0.33]{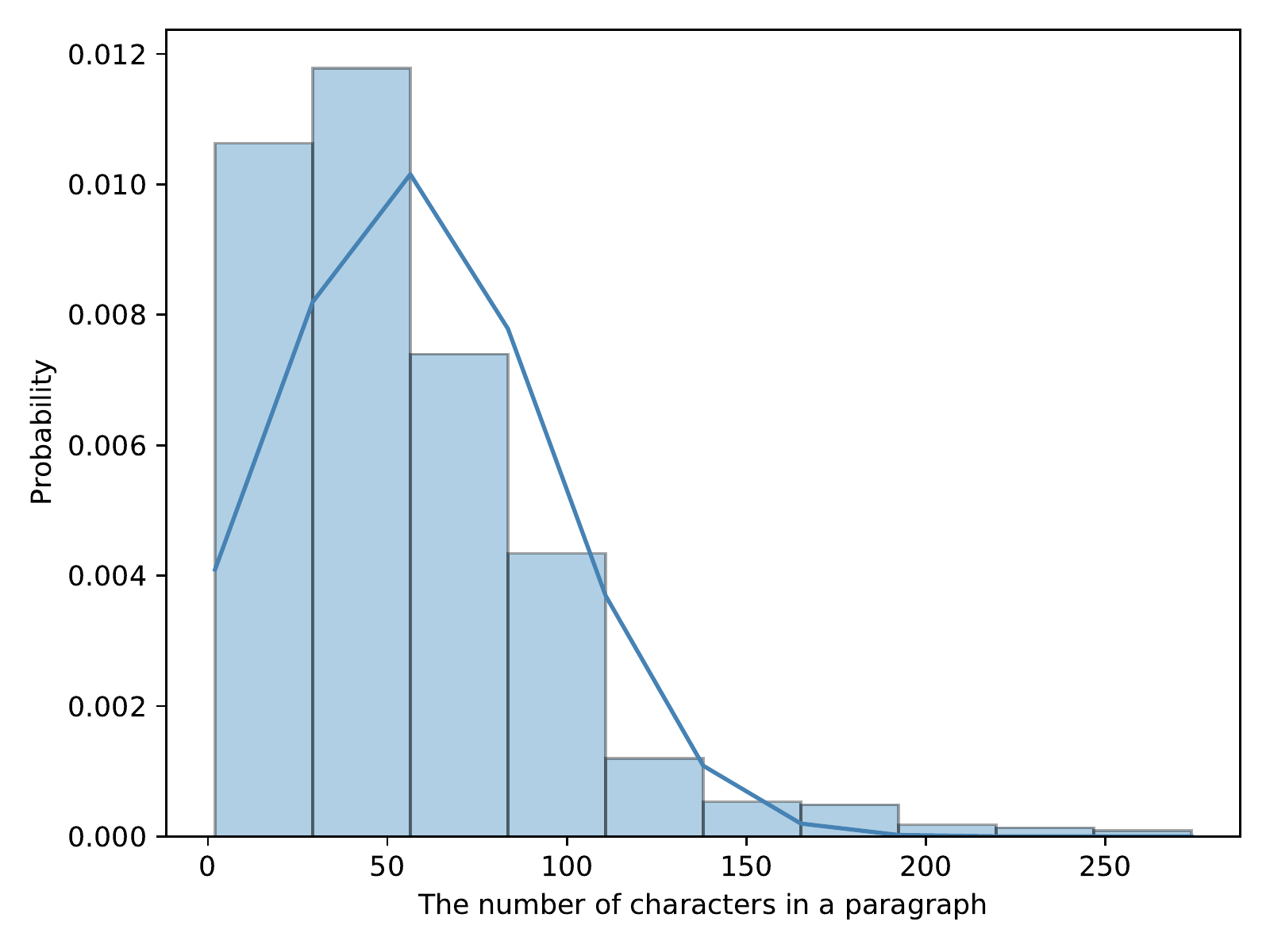}}
		\end{minipage}}
		\vfill
		\subfigure[The distribution of sentence durations. ]{
			\begin{minipage}{0.3\linewidth}
				\centerline{\includegraphics[scale=0.33]{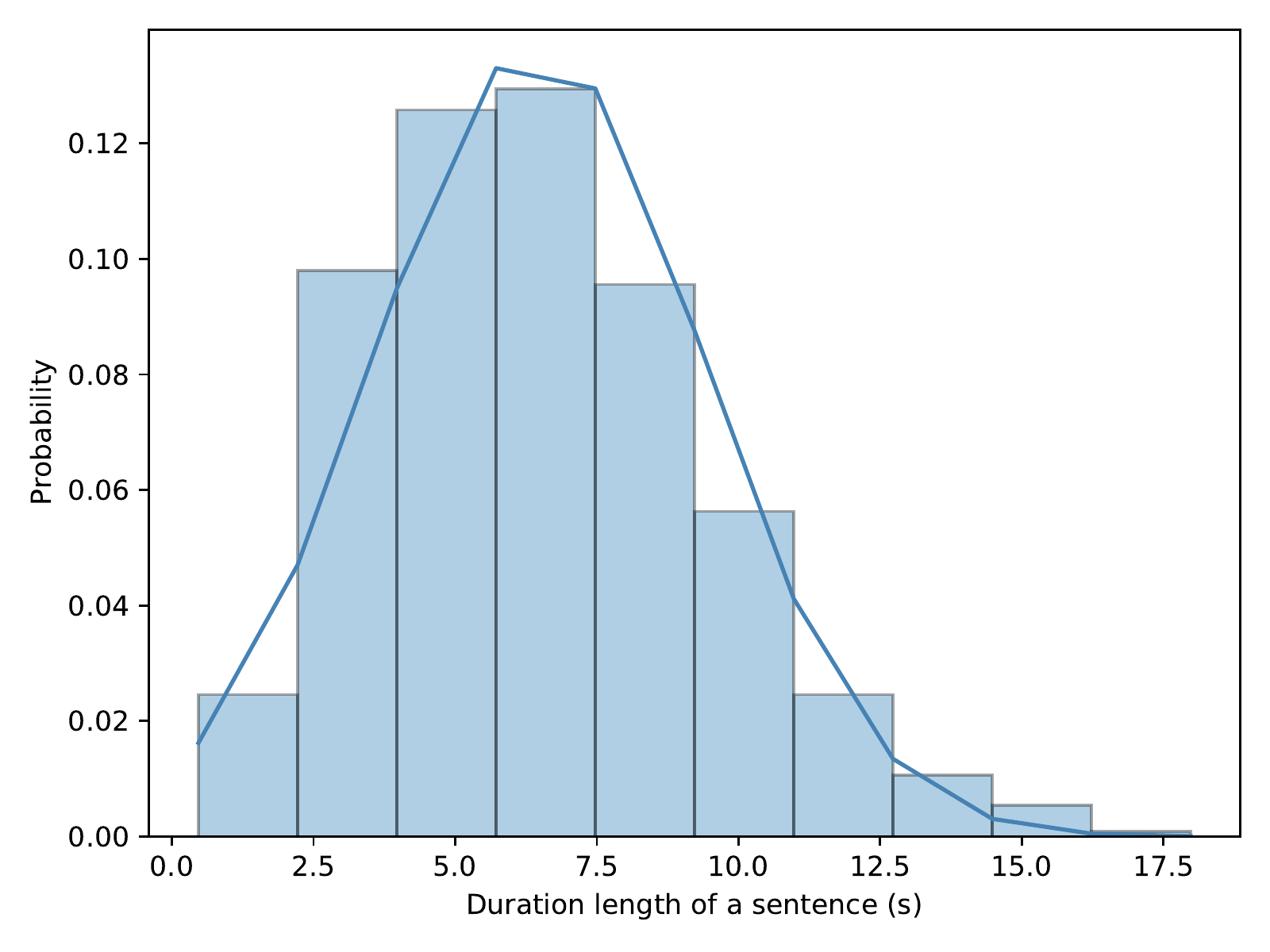}}
		\end{minipage}}
		\hfill		
		\subfigure[The distribution of paragraph durations. ]{
			\begin{minipage}{0.3\linewidth}
				\centerline{\includegraphics[scale=0.33]{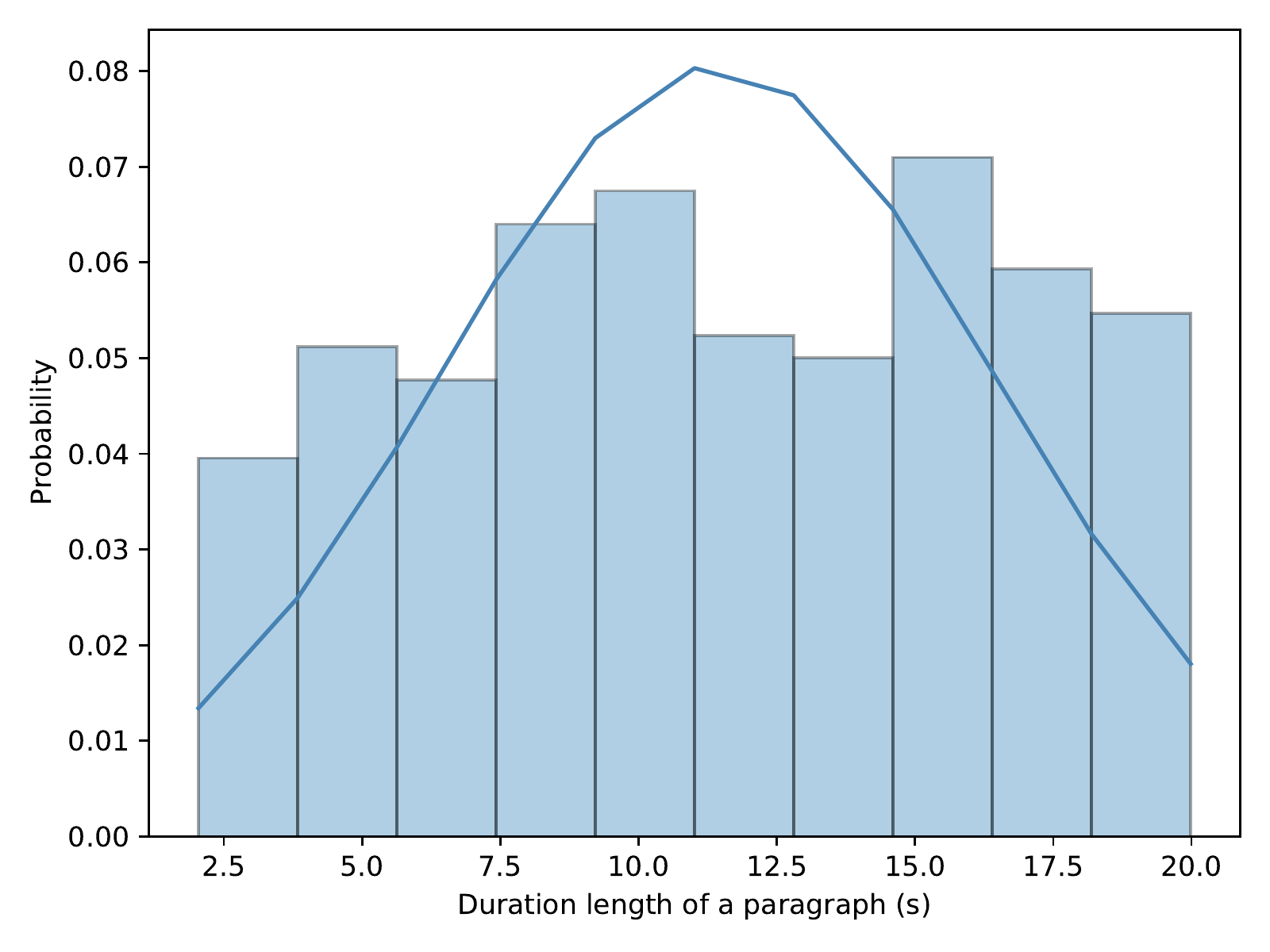}}
		\end{minipage}}

		\caption{Statistics of the training corpus. (a) The distribution of the number of sentences in a paragraph. (b) The distribution of the number of Chinese characters in a sentence. (c) The distribution of the number of Chinese characters in a paragraph. (d) The distribution of sentence durations. (e) The distribution of paragraph durations.}  \vspace*{-5pt}
		\label{fig:stats}
	\end{figure*}

	\begin{figure*}[h]
		\subfigure[LF0 variation curve]{
			\begin{minipage}{0.3\linewidth}
				\centerline{\includegraphics[scale=0.4]{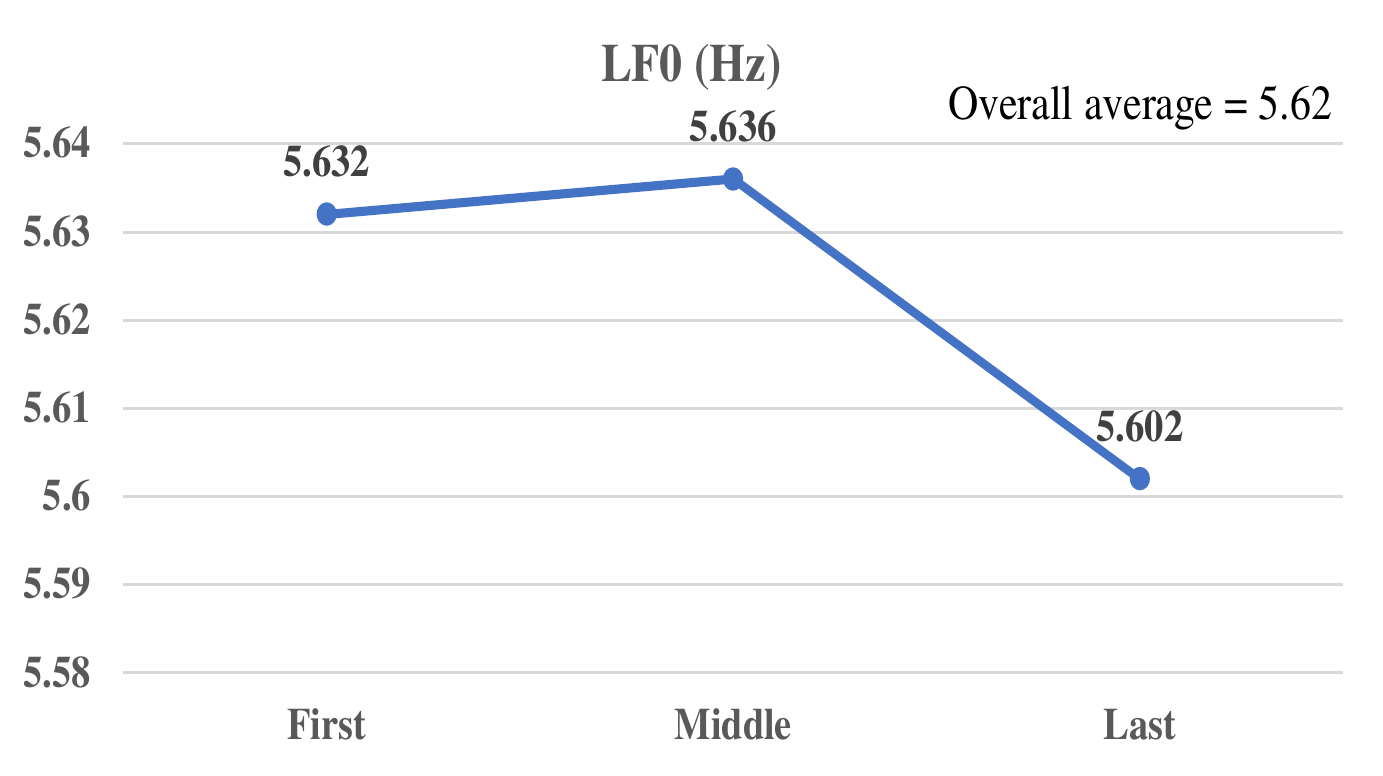}}
		\end{minipage}}
		\hfill
		\subfigure[Intensity variation curve]{
			\begin{minipage}{0.3\linewidth}
				\centerline{\includegraphics[scale=0.4]{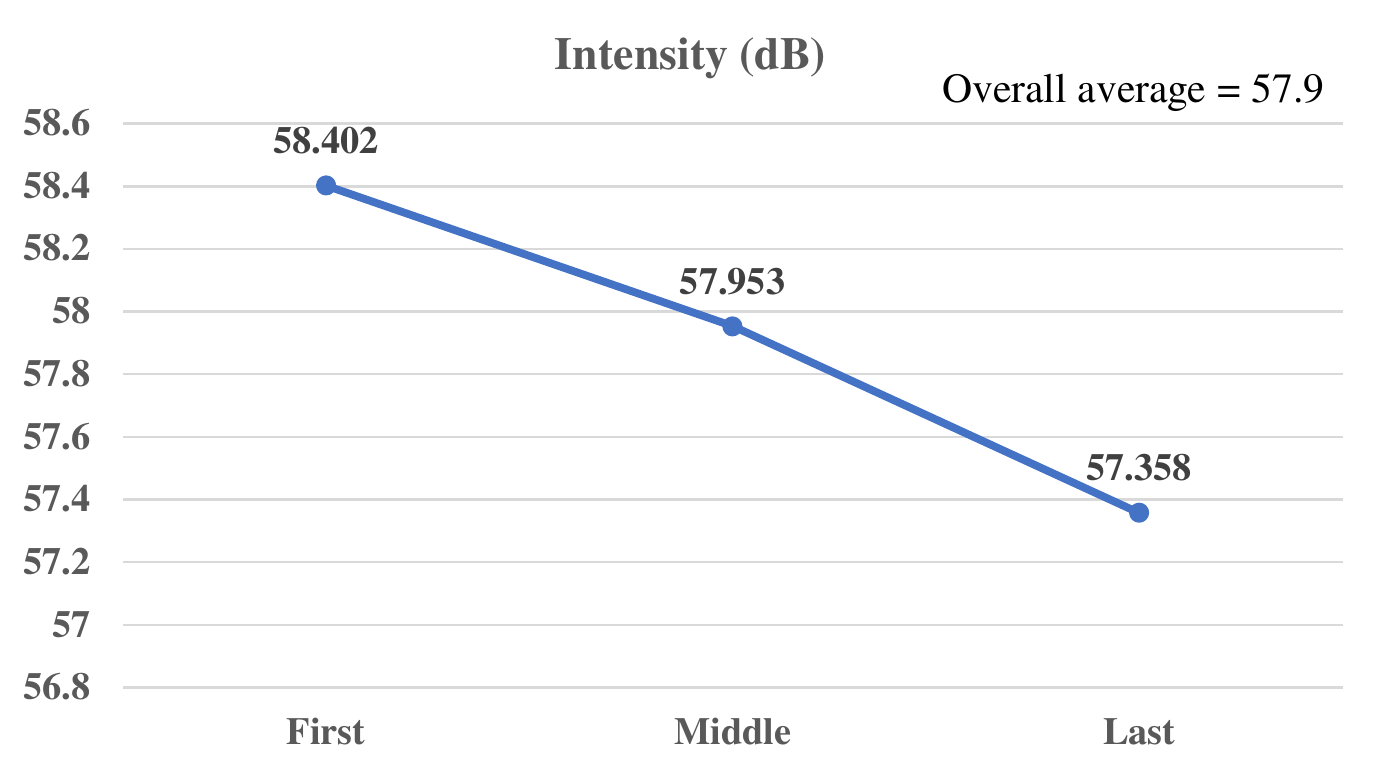}}
		\end{minipage}}
		\hfill
		\subfigure[Speech rate variation curve]{
			\begin{minipage}{0.3\linewidth}
				\centerline{\includegraphics[scale=0.4]{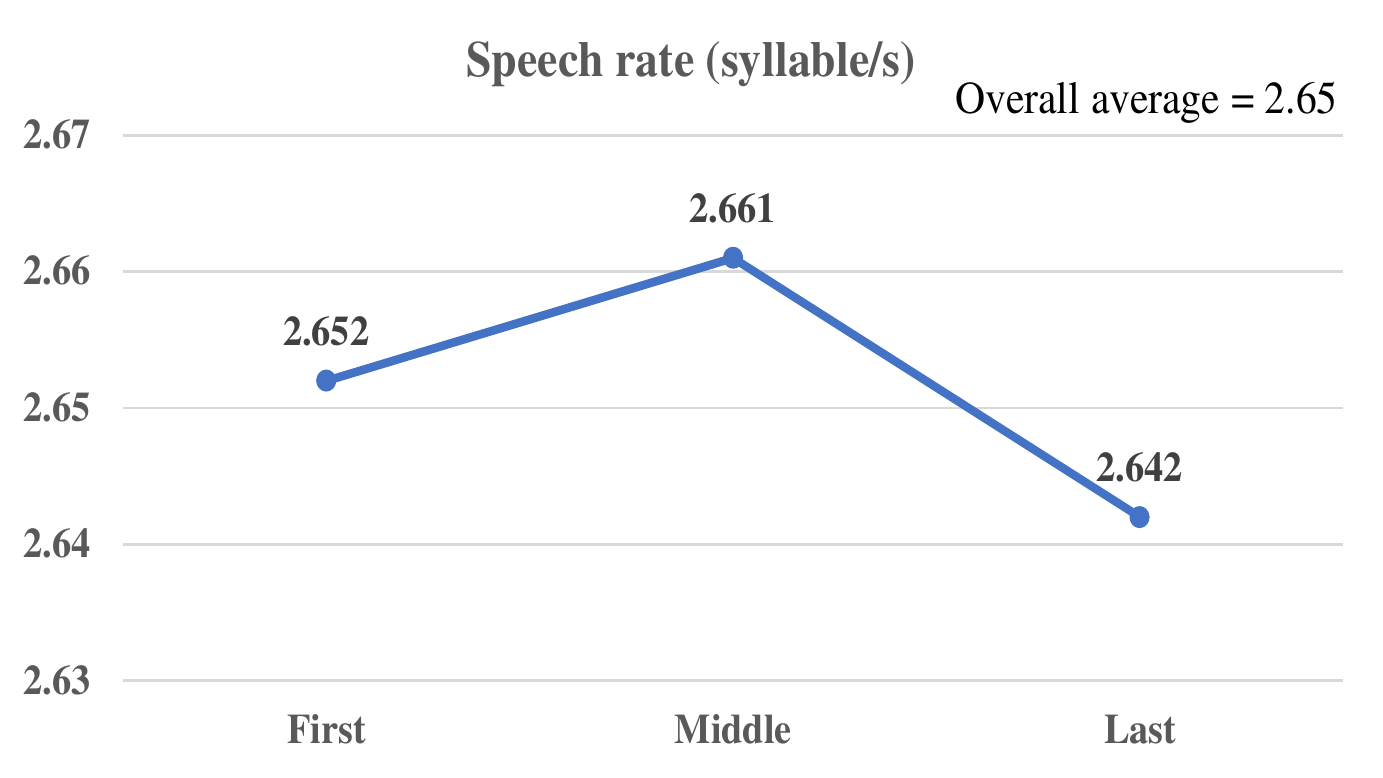}}
		\end{minipage}}
		\hfill
		\caption{Intra-paragraph prosody patterns analysis.}
		\label{fig:tendency}
	\end{figure*}
		
\subsection{The Sentence-position Network}
Distinctive differences in prosody can be found at the beginning, middle, and end sentences of a paragraph~\cite{Farrs2016ParagraphbasedPC}. We analyze the corresponding prosody patterns in paragraphs in section~\ref{sub:corpus}. To incorporate this prosodic information into the TTS model, we utilize a sentence-position network module which is composed of an up-sampling layer followed by a linear layer. The input to the sentence-position network is a 3-dimensional, one-hot code encoded as the first, middle or last sentence in the paragraph. The up-sampling layer up-samples the position code from sentence to phoneme level by replicating it. Finally, the linear layer is used to project the 3-dimensional, phone-level position code to the preset dimension of 256, facilitating the addition operation with the encoder output of the TTS backbone model.

An example of the sentence position code is illustrated in Fig.~\ref{fig:sent_pos}. Given a sentence, the sentence position in the paragraph can be determined and the corresponding phoneme sequence can also be obtained through the front-end module. The sentence position codes of the first, the last and the middle sentences are encoded as 0, 2 and 1, respectively. Then the sentence position is up-sampled from sentence level to phoneme level according to the corresponding phoneme sequence length.

The TTS backbone model, conditioned on (1) the sentence position code, and (2) the linguistic and prosodic information of the paragraph, generates the mel spectrograms of sentences from the phoneme sequences of sentences in training or the mel spectrograms of paragraphs from the phoneme sequences of the paragraphs in inference. The total loss L of the proposed model to be optimized is:
	\begin{equation}
		\label{equ:loss}
		L = \lambda_{1}{L}_{recon} + \lambda_{2}{L}_{stop} + \lambda_{3}{L}_{prosody}
	\end{equation}	
where ${L}_{recon}$ is the mel spectrogram reconstruction loss of mean square error (MSE), ${L}_{stop}$ is stop token loss of cross entropy, and ${L}_{prosody}$ is prosody prediction loss of MSE. $\lambda_{1}$, $\lambda_{2}$, $\lambda_{3}$ are the weights of the corresponding losses, respectively.

\section{Experiments}
\label{sec:experiments}
We first introduce the basic information of the corpus used in our experiments and then analyze the statistics of the corpus, intra-paragraph and inter-paragraph patterns. For experimental tests, we calculate objective metrics and also conduct subjective evaluations to measure the performance of the proposed model in generating paragraph speech.

	\begin{table}[h]
		\caption{The information of the training set and test set.}
		\label{tab:corpus}
		\centering
		\small
		\setlength{\tabcolsep}{3.5pt}
		
	    \begin{tabular}{|m{50pt}<{\centering}|m{33pt}<{\centering}|m{50pt}<{\centering}|m{43pt}<{\centering}|m{32pt}<{\centering}|} 	\hline		
			             & \multirow{2}*{\#Stories}    & \multirow{2}*{\#Paragraphs}        & \multirow{2}*{\#Sentences }     & \multirow{2}*{Hours}               \\ 
			                &           &                       &               & \\  \hline
			                
			Training set    & 40           & 801                 & 2,525          & 4.08           \\
			test set     & 4            & 32                  & 155            & 0.19                 \\
			Total           & 44           & 833                 & 2,680           & 4.27                 \\
			\hline
		\end{tabular}
	\end{table}

\subsection{Corpus Information and Analysis}
\label{sub:corpus}
\textbf{Basic information} In this work, we use a fairy-tale audio-book corpus to train and evaluate the proposed model. The information of the corpus is listed in Table~\ref{tab:corpus}. The corpus contains 44 stories recorded by a Chinese female mimicking children's voices, about 4.27 hours in total. We randomly select 40 stories as the training set and split the stories into 801 paragraphs and 2,525 utterances, about 4.08 hours. The rest of the 4 stories are used as the test set, which is split into 32 paragraphs. 

\textbf{Statistical analysis} Then text length (in terms of sentences and Chinese characters) and speech duration (in seconds) statistics are presented in Fig.~\ref{fig:stats}, in which (a), (b), and (c) are the distributions of the number of sentences in a paragraph, the number of Chinese characters in a sentence and a paragraph, respectively. On the average, each paragraph has 3 sentences, 55 Chinese characters, and each sentence has 17 Chinese characters. Additionally, (d) and (e) present the distributions of the duration length of sentences and paragraphs. The average duration lengths of a sentence and a paragraph are 6.5 and 11.4 seconds, respectively\footnote{The average number of sentences in a paragraph is 3, but the mode, referring to the value that occurs with the highest frequency is 2 due to the asymmetrical nature of the distribution. Consequently, the average duration length of a paragraph is 11.4 seconds, which is about twice as long as that of a sentence (6.5 seconds).}.

\textbf{Intra-paragraph prosody patterns analysis} To understand the variation of prosody features within a paragraph, we perform a statistical analysis of the individual prosodic features in sentences in the first, middle and last positions in a paragraph. Specifically, we calculate the mean values of sentence-level prosody features in different positions and plot the variation curves as exemplified in Fig.~\ref{fig:tendency}, where two prosody patterns are observed. \textbf{Declination}: pitch and intensity decline along with the paragraph. \textbf{Lengthening}: speech rate is faster in the middle than in the initial and final position of a paragraph, lengthening in the initial and final position. We observe that the range of prosodic features among three different sentence positions is not big, which may be attributed to two reasons: the scale of the corpus is relatively small, and the speaking style is not very distinctively different.

	\begin{table}[h]
		\caption{Inter-paragraph prosody patterns analysis.}
		\label{tab:inter_parp_pattern}
		\centering
		\small
		\setlength{\tabcolsep}{3.5pt}
		\begin{tabular}{|p{130pt}<{\centering}|p{40pt}<{\centering}|p{40pt}<{\centering}|}
			\hline		
			\multirow{2}*{Prosody features}           & \multirow{2}*{No break}   &  \multirow{2}*{Break}               \\
			& & \\ \hline
			LF0 diff (HZ)                   & 0.011                     & -0.021                    \\
			Intensity diff (dB)             & 0.368                     & -0.962                            \\
			Speech rate diff (syllable/s)   & 0.003                     & -0.003                            \\
			\hline
		\end{tabular}
	\end{table}

\begin{table}[h]
\newcommand{\tabincell}[2]{\begin{tabular}{@{}#1@{}}#2\end{tabular}}
	\caption{ Hyper-parameters of the proposed model. "conv1d-k-c-ReLU" denotes 1-D convolution with width k and c output channels with ReLU activation. "conv2d-(k, k)-(s, s)-c-BN-ReLU" denotes 2-D convolution with kernel size (k, k), stride (s, s) and c output channels with batch normalization and ReLU activation. FC-i-sigmoid stands for fully-connected layer with i units and sigmoid activation. LN means linear normalization.} 

 \centering
	\label{tab:hyparam}  
	\centering
	\begin{tabular}{l|l}
		\midrule
		 {Phone embedding} &{\tabincell{l}{ 512-D  } }      \\	\cline{1-2} 
		 {\makecell[c]{Text encoder}} &{\tabincell{l}{
		                                Pre-net: FC-512-ReLU $\rightarrow$ Dropout(0.1) \\  
		                                \quad $\rightarrow$ FC-512-ReLU  $\rightarrow$ Dropout(0.1) \\
		                                CBHG: Conv1D bank:  K=16, conv1d-k-256-ReLU \\
		                               \quad  Max pooling: stride=1, wide=2 \\
		                              \quad Conv1D projections: conv1d-3-256-ReLU \\ 
		                              \quad \quad  $\rightarrow$  conv1d-3-256-Linear \\
		                              \quad  Highway net: 4 layers of  FC-128-ReLU \\
		                              \quad  Bidirectional GRU: 128 cells \\
		                           } }      \\	\cline{1-2} 	
		 {\makecell[c]{Linguistics-aware\\ network}} &{\tabincell{l}{Paragraph text encoder: \\
		                                \quad Pre-net: 2 layers of FC-512-ReLU \\
		                                 \quad \quad $\rightarrow$ Dropout(0.1) \\
		                               \quad CBHG: Conv1D bank: \\ 
		                               \quad \quad  K=16, conv1d-k-256-ReLU \\
		                                    \quad \quad  Max pooling: stride=1, wide=2 \\
		                                    \quad \quad  Conv1D projections: conv1d-3-256-ReLU \\ 
		                                    \quad  \quad \quad  $\rightarrow$  conv1d-3-256-Linear \\
		                                    \quad  \quad  Highway net: 4 layers of FC-128-ReLU \\
		                                    \quad  \quad Bidirectional GRU: 128 cells \\
		                              Multi-head attention: \\
		                              \quad Attention heads: 4 \\
		                              \quad Attention dimension: 256-D\\
		                              
		                               } }      \\	\cline{1-2} 
		 {\makecell[c]{Prosody-aware \\ network}} &{\tabincell{l}{Paragraph prosody encoder: \\
		                                         \quad Conv2D bank: K=(32, 32, 64, 64, 128, 128), \\
		                                     \quad \quad \qquad\ \ \quad conv2d-(3,3)-(2,2)-k-BN-ReLU \\
		                                     \quad   GRU: 256 cells  \\

            		                              Multi-head attention: \\
            		                              \quad Attention heads: 4 \\
            		                              \quad Attention dimension: 256-D\\
		                                         Paragraph prosody predictor: \\
		                                         \quad Conv1D projections: conv1d-3-256 $\rightarrow$ LN \\
		                                         \quad \qquad\ \ \quad  $\rightarrow$  Dropout(0.1)  $\rightarrow$ conv1d-3-256  \\
		                                         \quad \qquad\ \ \quad $\rightarrow$ LN $\rightarrow$  Dropout(0.1) \\
		                                         \quad FC: FC-3 \\

		                               } }      \\	\cline{1-2} 	
		 {\makecell[c]{Sentence-position\\ network}} &{\tabincell{l}{FC: FC-256 \\} }      \\	\cline{1-2} 
		 {GMM attention} &{\tabincell{l}{ Mixtures: 8 \\
		                                  Attention dimension: 128-D\\
		                                } }      \\	\cline{1-2} 	
		                                
		 {\makecell[c]{Decoder}}  & {\tabincell{l}{Pre-net: FC-256-ReLU  $\rightarrow$ Dropout(0.2) \\
		                             \qquad\ \ \quad $\rightarrow$ FC-256-ReLU  $\rightarrow$ Dropout(0.2) \\
		                             Decoder RNN: 2-layer LSTM (1024 cells)\\
		                             Post-net:  4 layers of conv1d-5-512-BN-ReLU \\
		                  \qquad\ \ \quad $\rightarrow$ conv1d-5-512-BN    \\
		                             Linear projection: FC-80\\
		                             Stop token projection: FC-1-sigmoid}}  \\ 

		                             
		\bottomrule
	\end{tabular}
\end{table}

	\begin{table*}[h]
		\caption{Objective test results for MCD, syllable-level prosody correlation and pause RMSE.}
		\label{tab:obj_test}
		\centering
		\small
		\setlength{\tabcolsep}{3.5pt}
		\begin{tabular}{|p{115pt}<{}|p{40pt}<{}|p{130pt}<{}|p{130pt}<{}|p{60pt}<{}|}
			\hline		
			\multirow{2}*{\qquad Models}     & \multirow{2}*{\quad MCD $\downarrow$ }                    & \multicolumn{2}{c|}{Syllable-level prosody correlation $\uparrow$ }                            &  \multirow{2}*{Pause RMSE  $\downarrow$ }     \\ \cline{3-4} 
			                         &                          &\qquad LF0                                      &\qquad Duration                                      & \\ \hline
		\qquad	Baseline                   & \quad 4.524               & \qquad 0.708 \  (p-value=1.96e-05)                 & \qquad 0.739 \ (p-value=0.02)              & \qquad 0.499                    \\
		\qquad	LingTTS                    & \quad 4.434               & \qquad 0.720 \ (p-value=8.52e-06)                  &  \qquad 0.762 \  (p-value=6.83e-04)                &  \qquad 0.404                            \\
		\qquad	ProsTTS                    & \quad 4.430               & \qquad 0.714 \ (p-value=1.59e-06)                 & \qquad 0.735 \  (p-value=0.01)                  &\qquad 0.392                            \\
		\qquad	ComTTS                     & \quad 4.437               & \qquad 0.721 \ (p-value=2.45e-06)                 & \qquad  0.747 \  (p-value=5.65e-03)                 &\qquad \textbf{0.380}                   \\
		\qquad	ParaTTS                    & \quad \textbf{4.351}      & \qquad \textbf{0.723} \   (p-value=4.51e-06)       & \qquad  \textbf{0.770} \  (p-value=4.44-e04)                 & \qquad 0.415        \\ \hline 
		\qquad	ParaTTS (GT\_prosody)      & \quad 4.308      & \qquad 0.723 \   (p-value=3.93e-05)       & \qquad  0.780 \  (p-value=0.01)                 &\qquad 0.401                          \\  \hline
		\end{tabular}
	\end{table*}

\textbf{Inter-paragraph prosody patterns analysis} We also analyze the prosody feature variation across paragraph boundaries (break) and within a paragraph (no break), as shown in Table~\ref{tab:inter_parp_pattern}. The values in the break are the mean discrepancy of prosody features between the last sentence in the current paragraph and the first sentence in the next paragraph. And the values in no break are the mean discrepancy of prosody features between the current sentence and the succeeding sentence within a paragraph. The positive values in no break suggest that there is a declination of prosody attributes within a paragraph. While, the negative values in break indicate that the phenomenon of \textbf{prosody reset} appears in paragraph boundaries, which means the LF0, intensity and speech rate increase at the beginning of a new paragraph. In this work, we focus on individual paragraph speech synthesis so that the inter-paragraph prosody patterns are not considered. 

\subsection{Experimental Setup}
The relatively small size of the training data used in this study is a challenge to training a highly stable, end-to-end TTS model. Thus, we firstly pre-train the backbone of the modified Tacotron2 using a standard TTS corpus, containing 17.83-hour reading-style Chinese female speech data. Then we fine-tune the model for ParaTTS using the audio-book corpus. In the training stage, sentence phoneme sequences are fed into the model as input. Mel spectrograms are extracted from recordings, which are down-sampled from 44.1 kHz to 16kHz, and used as the target output. To obtain phone-level prosody features in paragraph prosody extractor, we first extract frame-level LF0 and intensity values using the Python library of Parselmouth\footnote{Parselmouth provides a complete and Pythonic interface to the internal Praat code that can be used for speech analysis of pitch and intensity. Parselmouth can be found at\url{ https://github.com/YannickJadoul/Parselmouth} and Praat can be found at \url{https://www.fon.hum.uva.nl/praat/} }. Meanwhile, we conduct force alignment using Hidden Markov Model Toolkit (HTK)~\cite{young2002htk} to get the phone duration, and then calculate the phone-level LF0 and energy by averaging frame-level values based on the frame lengths of each phone. We train models on a single GPU with batch size of 16 up to 400k steps for the pre-trained model and 200k steps for the fine-tuned models, using Adam optimizer~\cite{kingma2014adam} with $\beta_{1}$ = 0.9 and $\beta_{2}$ = 0.999. The hyper-parameters of the model used in our experiments are described in Table~\ref{tab:hyparam}. At the inference stage, paragraph phoneme sequence is fed into the model as input. The output mel spectrogram is transformed into a waveform using the multi-band WaveRNN vocoder~\cite{Yu2020DurIANDI}, which is pre-trained to 500k steps using the standard corpus and adapted to 200k steps using the audio-book corpus.

In our evaluations, we compare the following five models for paragraph-level speech synthesis.

	\begin{itemize}
		\item \textbf{Baseline}: the modified Tacotron2 as described in Section~\ref{subsec:taco2}.
		\item \textbf{LingTTS}: Baseline with the linguistics-aware network.
		\item \textbf{ProsTTS}: Baseline with the prosody-aware network.
		\item \textbf{ComTTS}: Baseline with the combination of the linguistics-aware and prosody-aware networks.
		\item \textbf{ParaTTS}: ComTTS with the sentence-position network.
	\end{itemize}
	
\subsection{Objective Evaluation}
\textbf{Naturalness}
We calculate mel-cepstrum distortion (MCD) to measure the naturalness objectively. Before computing MCD, we use dynamic time warping to align predicted and target mel spectrogram sequences because the lengths of the two sequences can be different. The MCD result is shown in the second column in Table~\ref{tab:obj_test}. The model with linguistics-aware network (\textbf{LingTTS}) or prosody-aware network (\textbf{ProsTTS}) or both (\textbf{ComTTS}) gets a similar MCD and outperforms the \textbf{Baseline}. With the sentence-position network, the model (\textbf{ParaTTS}) decreases MCD further and achieves the lowest MCD.

\begin{figure*}[h]
    \begin{center}
     \includegraphics[scale=0.55]{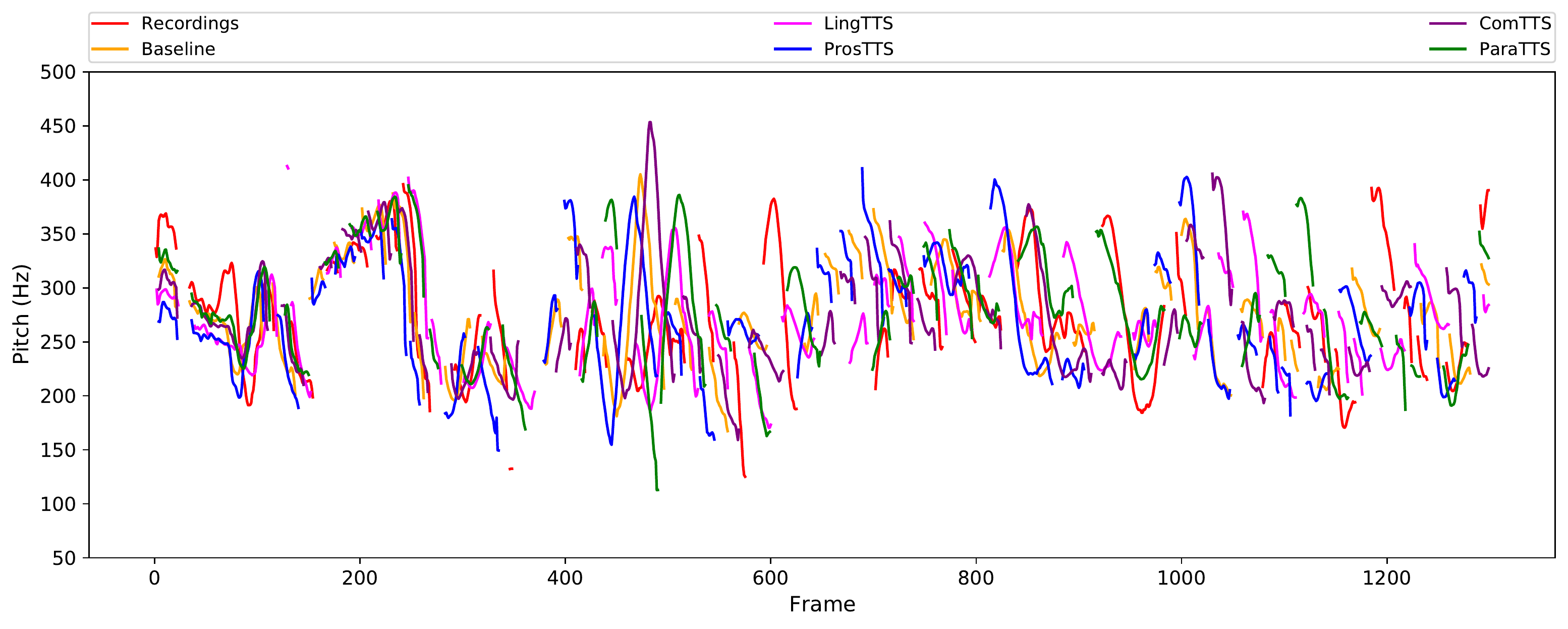}
    \end{center}  \vspace*{-10pt}
    \caption{The pitch contours of a paragraph from the recording and the synthesized results of the \textbf{Baseline}, \textbf{LingTTS}, \textbf{ProsTTS}, \textbf{ComTTS} and \textbf{ParaTTS}.}
    \label{fig:pitch_contour}	
\end{figure*}

\textbf{Prosody}
 To measure the synthesized prosody in a paragraph, we calculate the Pearson correlation coefficient~\cite{benesty2009pearson} on the prosody features at the syllable level, including LF0, intensity and duration. The LF0 and duration correlation results and the corresponding p-values are presented in Table~\ref{tab:obj_test}. We do not list the intensity correlation because all models have a comparable high correlation value, around 0.90. Regarding the LF0 and duration correlations, all models achieve good correlations, and \textbf{ParaTTS} is the best. From the results, it is observed that even though \textbf{ProsTTS} is more related to the prosody, it does not achieve better results than the \textbf{LingTTS}. This may be because the prosody prediction heavily dependents on the linguistic information. Consequently, \textbf{ComTTS} which combines the linguistic-aware network and prosody-aware network does not achieve additional benefits compared with \textbf{LingTTS}. After adding sentence-position network, \textbf{ParaTTS} obtains improved performance, indicating the benefits of the sentence position information, which explicitly provides the simple yet effective features of individual sentences in multi-sentence paragraph generation.
 
 We also generate the results by feeding the ground-truth prosody features to the \textbf{ParaTTS}, referred as \textbf{ParaTTS (GT\_prosody)}, to show the upper bound of the proposed model. The objective evaluations of \textbf{ParaTTS (GT\_prosody)} are presented in Table~\ref{tab:obj_test} as well. Obviously, \textbf{ParaTTS (GT\_prosody)} achieves the best results. Additionally, we also observe that \textbf{ParaTTS (GT\_prosody)} has a similar LF0 correlation with \textbf{ParaTTS} but obtains better MCD. The results indicate that the ground-truth prosody features benefit the naturalness and the prosody prediction in the proposed model is appropriate. To visualize the prosody prediction performance, we plot the pitch contours of a paragraph from the recording and the synthesized results of the \textbf{Baseline}, \textbf{LingTTS}, \textbf{ProsTTS} \textbf{ComTTS} and \textbf{ParaTTS}, as shown in Fig.~\ref{fig:pitch_contour}. We can observe that the pitch contour of \textbf{ParaTTS} is the closest to that of the recordings compared to the other models.

		\begin{figure*}[h]
		\subfigure[Recording]{
			\begin{minipage}{0.32\linewidth}
				\centerline{\includegraphics[scale=0.32]{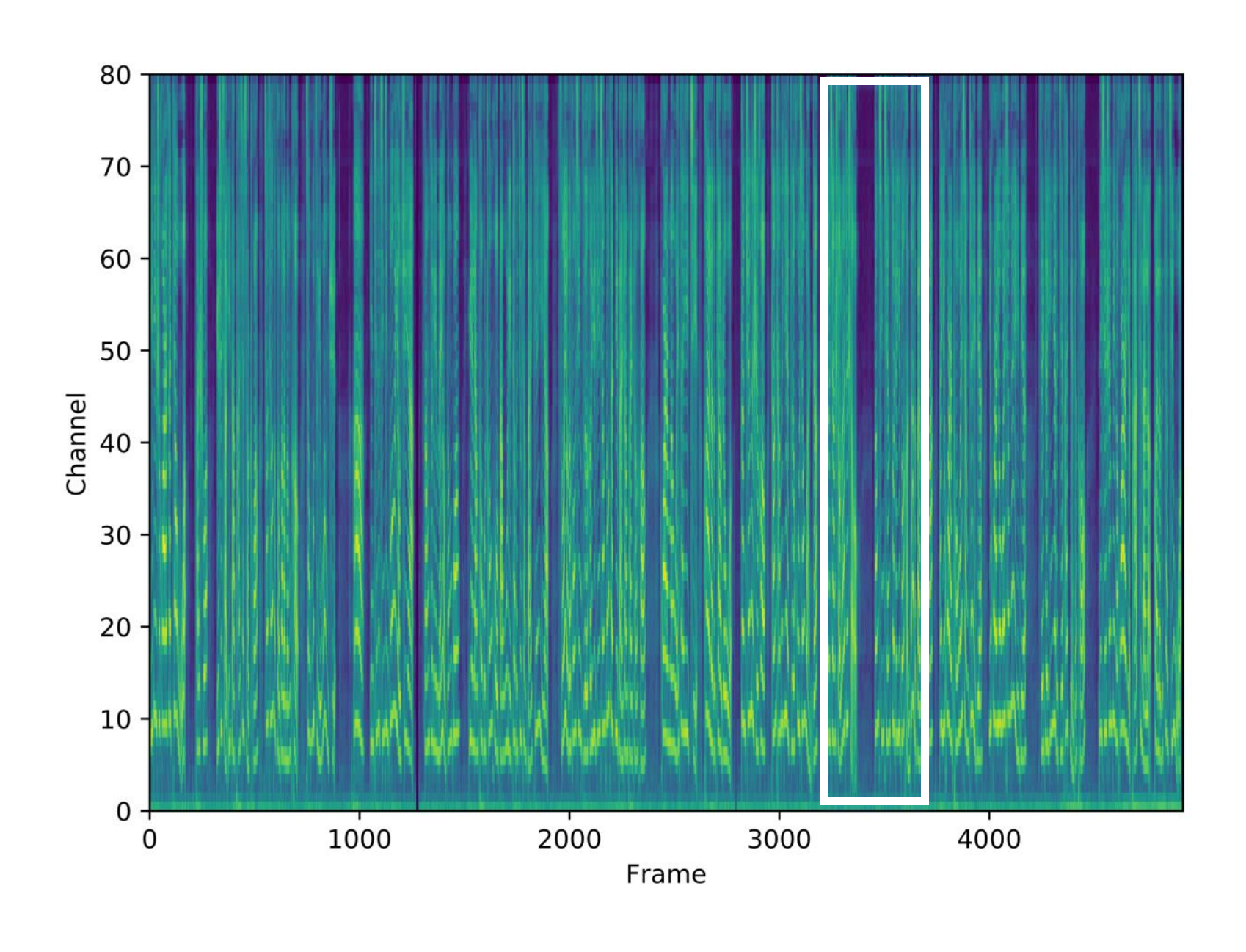}}
		\end{minipage}}
		\hfill
		\subfigure[Baseline]{
			\begin{minipage}{0.32\linewidth}
				\centerline{\includegraphics[scale=0.32]{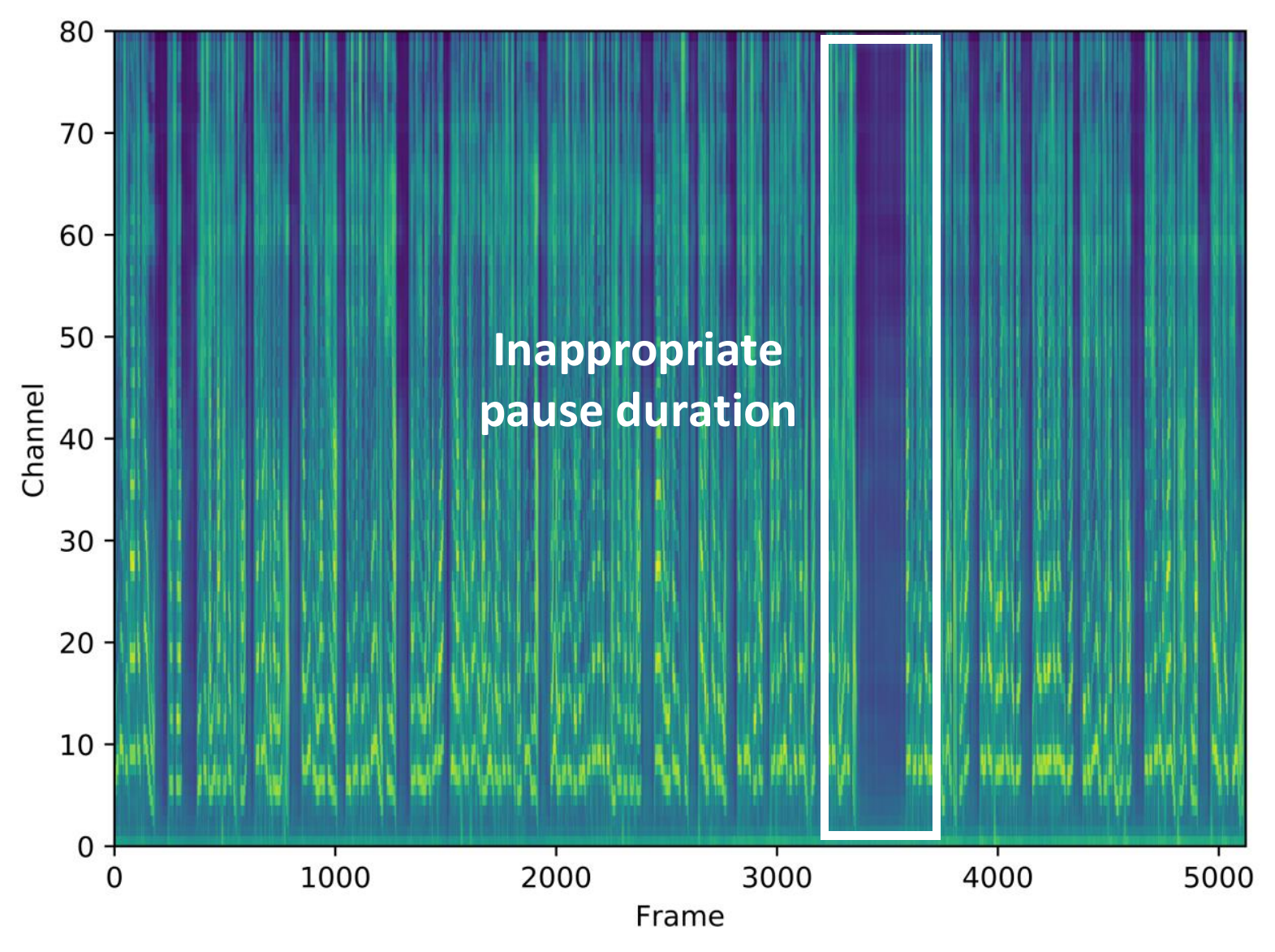}}
		\end{minipage}}
			\hfill
		\hfill
		\subfigure[LingTTS]{
			\begin{minipage}{0.32\linewidth}
				\centerline{\includegraphics[scale=0.32]{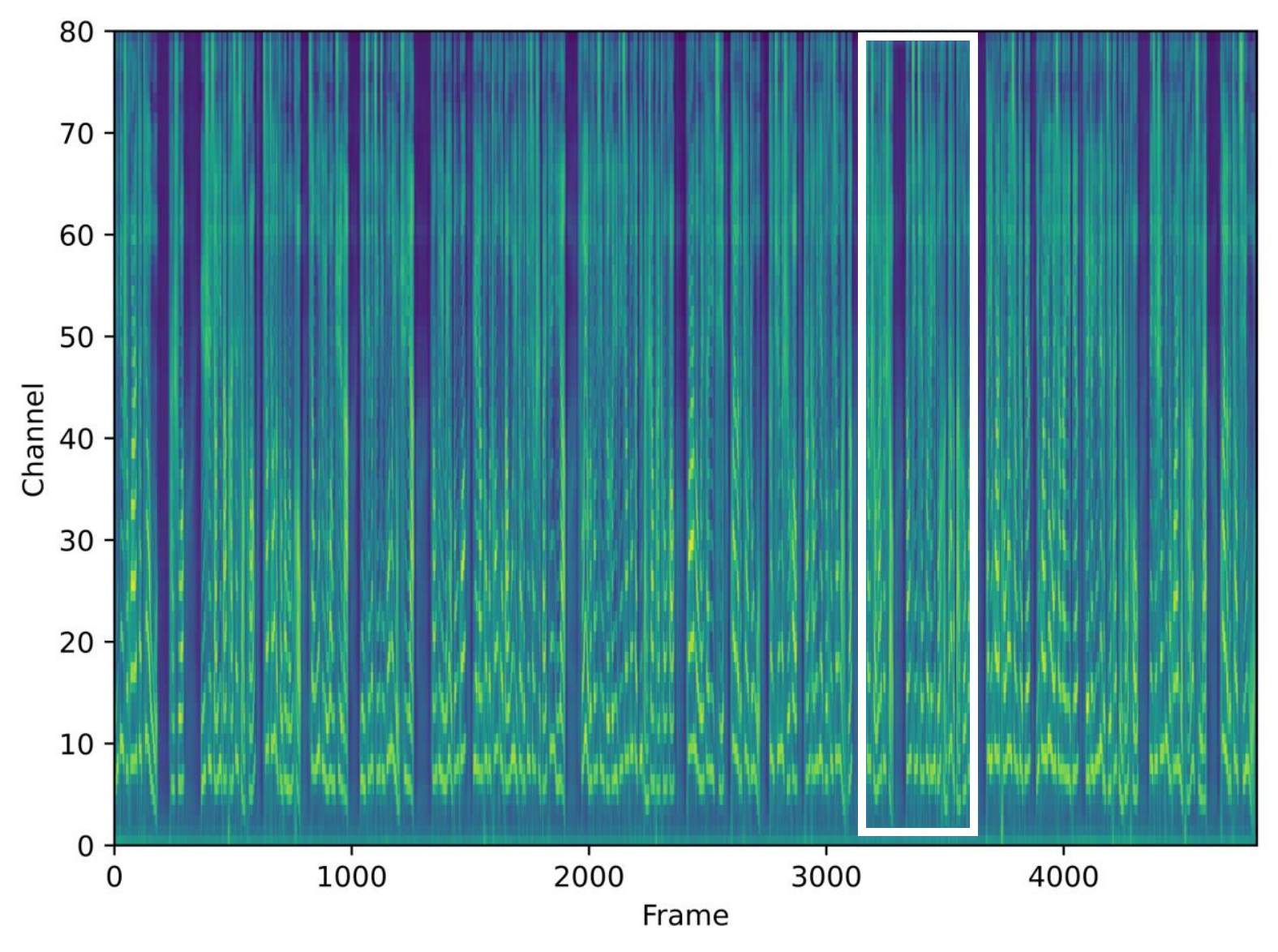}}
		\end{minipage}}
			\hfill
				\vfill
		\subfigure[ProsTTS]{
			\begin{minipage}{0.32\linewidth}
				\centerline{\includegraphics[scale=0.32]{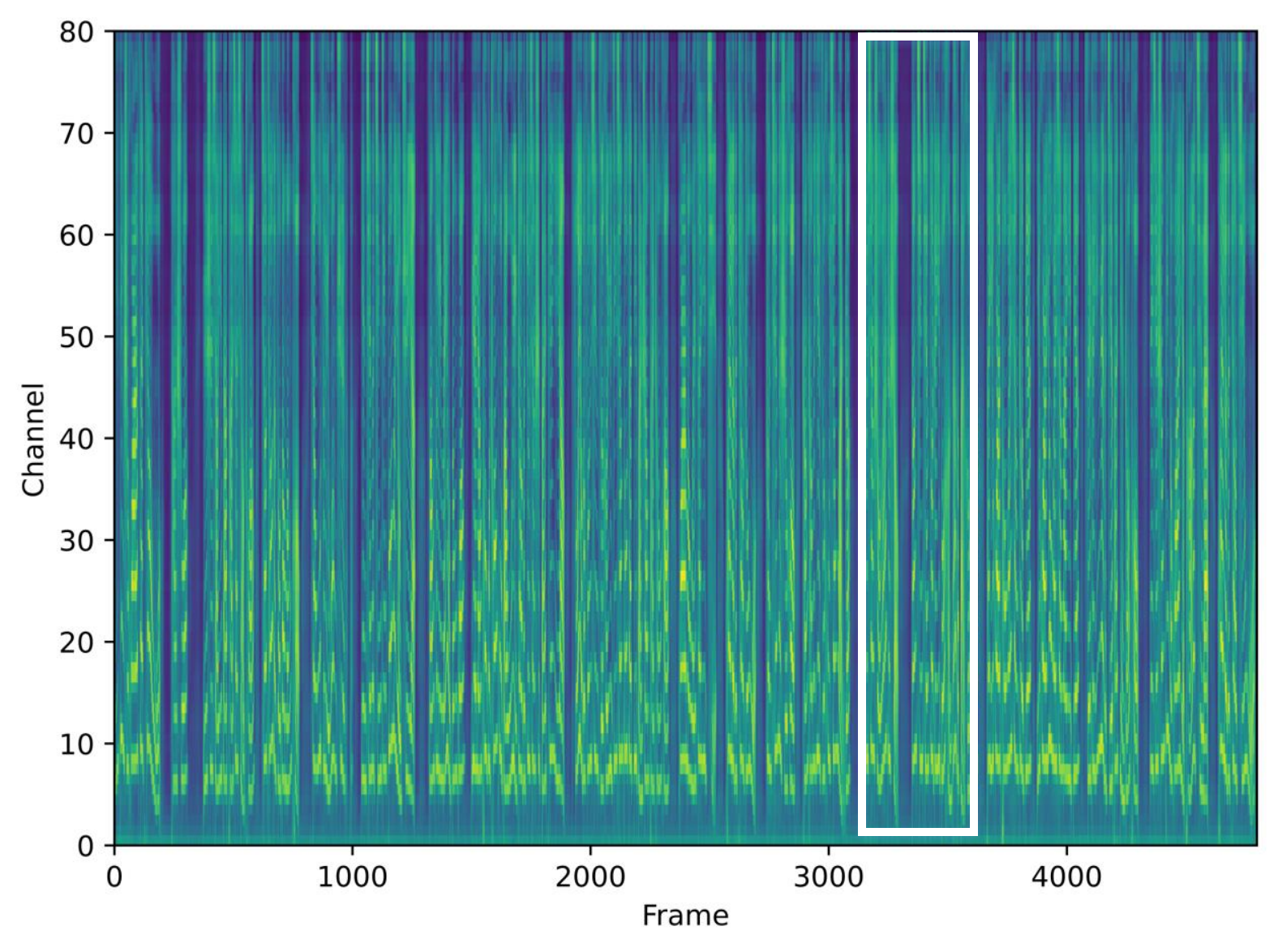}}
		\end{minipage}}
		\hfill
		\subfigure[ComTTS]{
			\begin{minipage}{0.32\linewidth}
				\centerline{\includegraphics[scale=0.32]{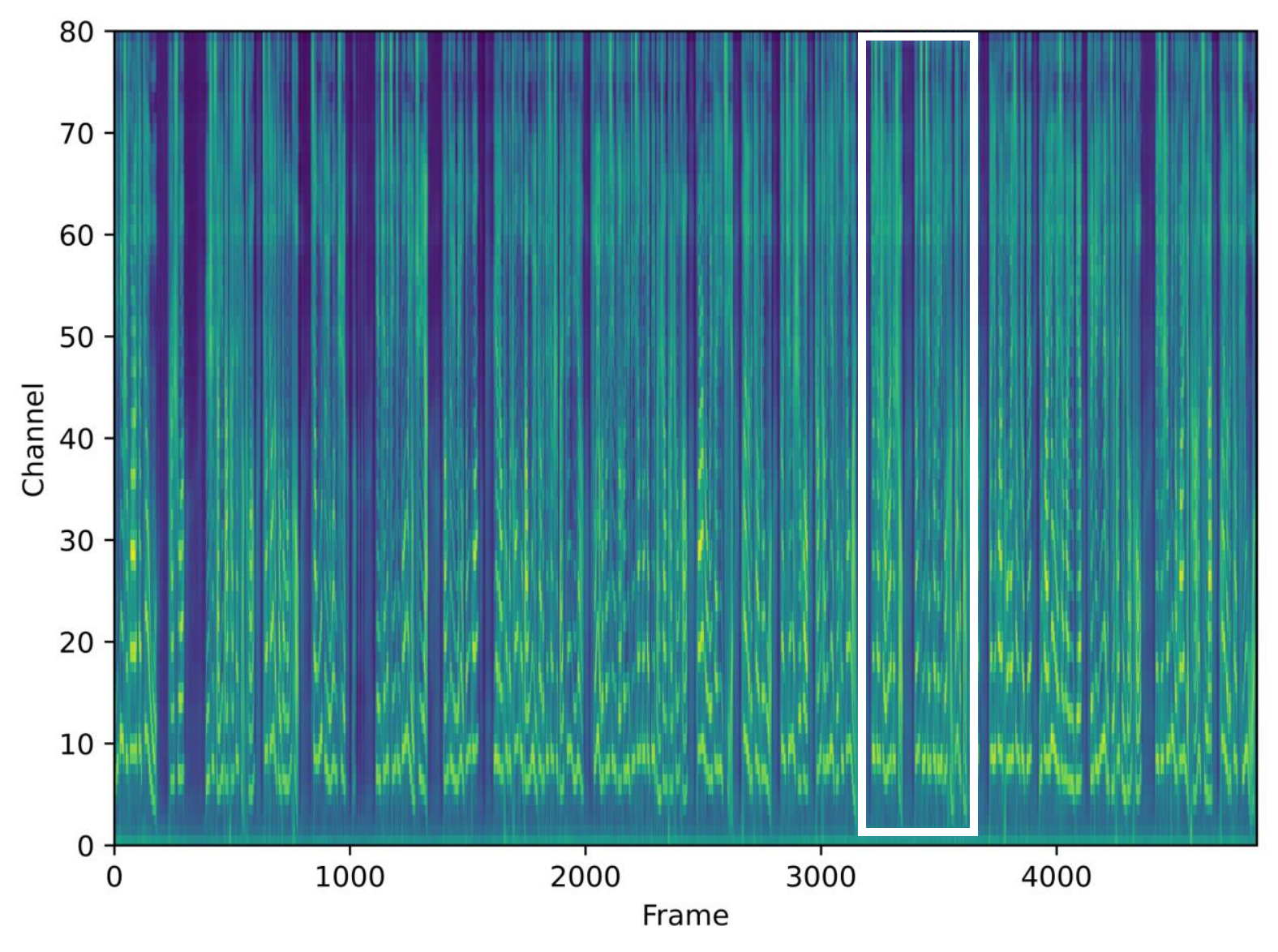}}
		\end{minipage}}
			\hfill
		\hfill
		\subfigure[ParaTTS]{
			\begin{minipage}{0.32\linewidth}
				\centerline{\includegraphics[scale=0.32]{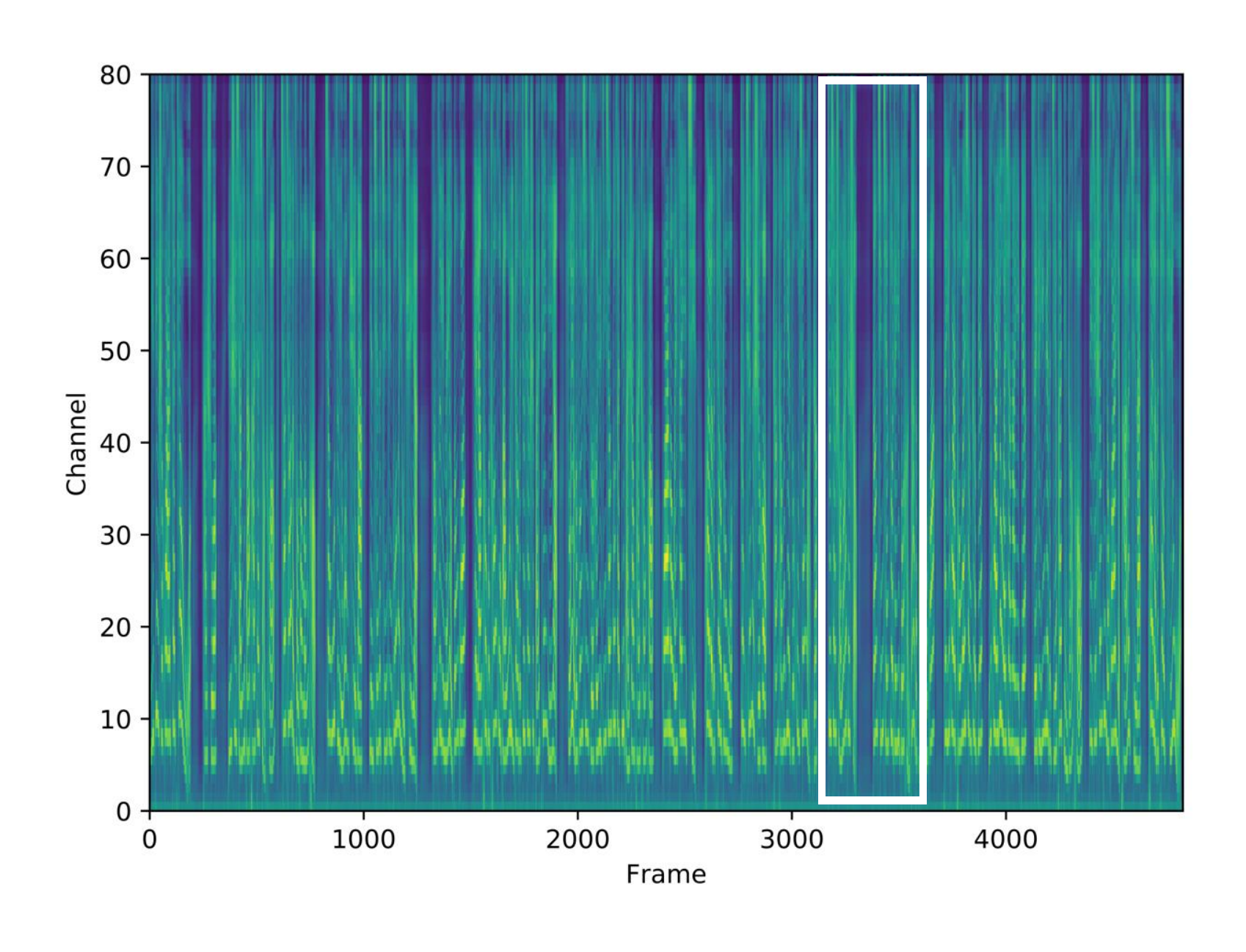}}
		\end{minipage}}
			\hfill
		\caption{Mel spectrograms extracted from the (a) recording, and the synthesized paragraph speech by the (b) \textbf{Baseline}, (c) \textbf{LingTTS}, (d) \textbf{ProsTTS}, (e) \textbf{ComTTS} and (f) \textbf{ParaTTS}. The white boxes present the pause duration between two sentences.} \vspace*{-5pt}
		\label{fig:melspec}
	\end{figure*}

\textbf{Break} 
It has been shown that pause duration is highly correlated to the discourse structure~\cite{Ouden2009ProsodicRO}. The pause duration, particularly between successive sentences, is used to introduce suspense and climax in storytelling, which can enhance the audience's attraction to the story and build some anticipation~\cite{Theune2006GeneratingES}. We calculate the root mean square error (RMSE) of pause duration between consecutive sentences in a paragraph to explore if the models can learn the break across sentences. To be specific, the pause duration between two sentences is the difference value between the end time of the last word in the current sentence and the start time of the first word in the next sentence. 

The pause duration RMSE result is shown in the last column in Table~\ref{tab:obj_test}. The \textbf{Baseline} achieves the worst RMSE result. The other four models all achieve better results than the \textbf{Baseline}, in which \textbf{ComTTS} is slightly better than others. We observe that \textbf{ParaTTS} can learn a more accurate pause duration than the \textbf{Baseline}. Fig.~\ref{fig:melspec} shows mel spectrograms of the recording and the synthesized paragraph speech by the \textbf{Baseline}, \textbf{LingTTS}, \textbf{ProsTTS}, \textbf{ComTTS} and \textbf{ParaTTS}, in which white boxes are the pause duration between two consecutive sentences. We can find that the pause duration in \textbf{Baseline} is too long to appropriate, which may cause unnatural perceived performance. The corresponding samples (numbered 1.3) can be found in our sound sample page\footnote{\url{https://lmxue.github.io/paratts/}}. We conjecture that the multi-head mechanisms between the sentence and multi-sentence paragraph in linguistics-aware and prosody-aware networks can learn the cross-sentence context in training, including the pause duration. 


\begin{figure}[h]
    \begin{center}
     \includegraphics[scale=0.6]{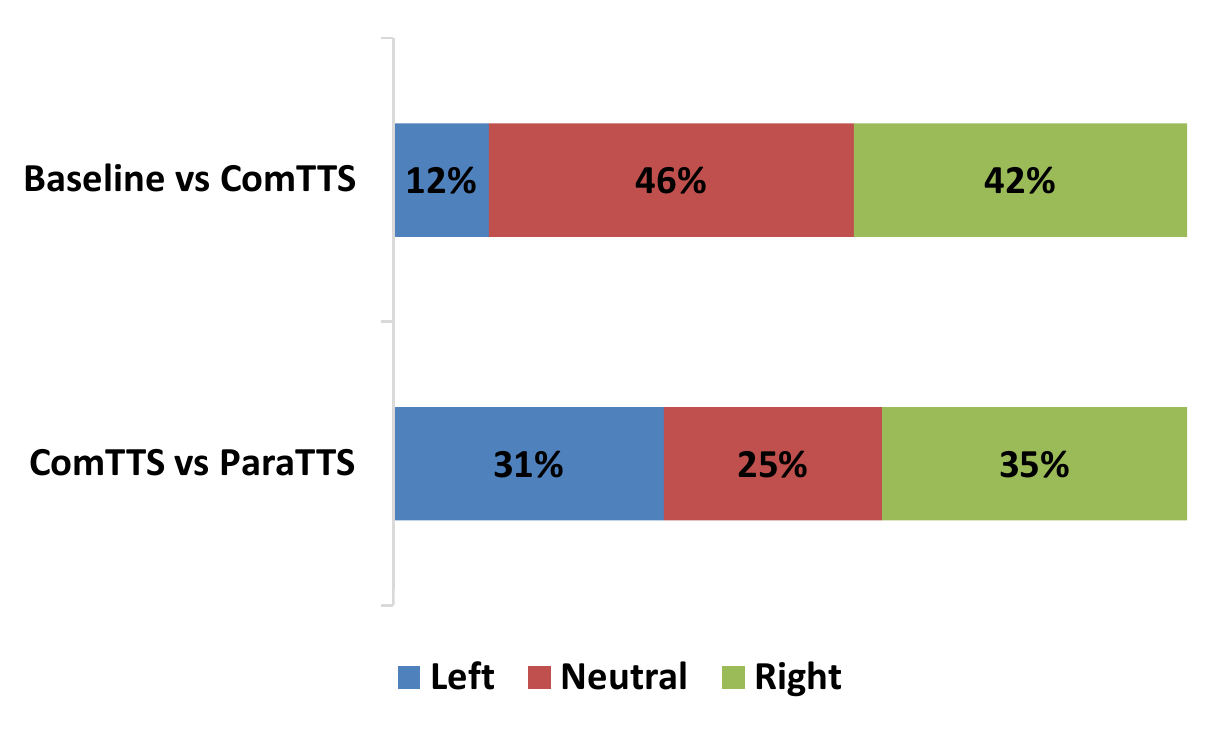}
    \end{center}  \vspace*{-10pt}
    \caption{Preference test results.}
    \label{fig:preference}	
\end{figure}

\subsection{Subjective Evaluation}
A group of 20 listening subjects who are native Chinese speakers with normal hearing participates in the subjective tests and evaluates each paragraph as a whole rather than in its isolated sentences, a similar testing was conducted in~\cite{Clark2019EvaluatingLT}.

\textbf{Preference test} The preference test between two models is to choose which model is preferable based upon the overall perceived impression. We first perform preference tests among \textbf{Baseline}, \textbf{ComTTS} and \textbf{ParaTTS} to compare the effectiveness of the linguistics-aware network, prosody-aware network and sentence-position network. The preference test results are presented in Fig.~\ref{fig:preference}. \textbf{ComTTS} gets 30\% more preference than the \textbf{Baseline}, indicating that the linguistic and prosodic information can indeed improve paragraph-based speech synthesis. With the additional sentence position information, \textbf{ParaTTS} gets an extra preference (4\%) over \textbf{ComTTS}. This small preference gain is also consistent with the slightly better MCD and prosody correlations. As described in the intra-paragraph prosody patterns analysis, the prosody variation range among three different sentence positions is not large, hence leading to a relatively smaller improvement. In the following subjective tests, we compare \textbf{Baseline}, \textbf{ParaTTS} and recordings with mean opinion score (MOS) scores in in-domain and out-of-domain tests.


\begin{figure}[h]
    \begin{center}
     \includegraphics[scale=0.65]{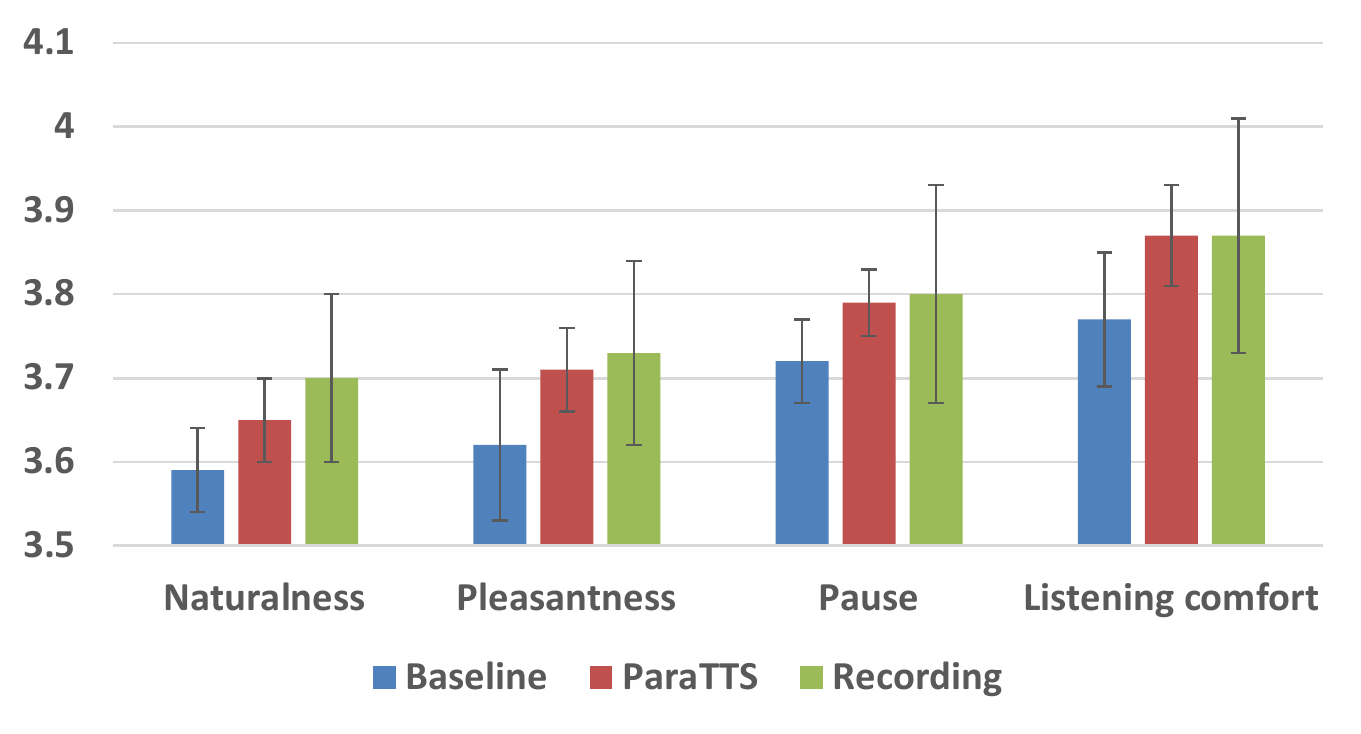}
    \end{center} \vspace*{-10pt}
    \caption{MOS scores with 95\% confidence intervals on four dimensions.}
    \label{fig:inpara_mos_4dims}	 \vspace*{-5pt}
\end{figure}

\textbf{Detailed MOS test}
To further evaluate the perceived quality of the synthesized paragraph speech, we conduct 5-point mean opinion score (MOS) tests in four different dimensions: naturalness, pleasantness, pause and listening comfort\footnote{The explanations for four criteria are:  naturalness: "very natural" to "unnatural"; pleasantness: "very unpleasant" to "very pleasant"; pause: “speech pauses confusing/unpleasant” to “speech pauses appropriate/pleasant”; listening comfort: “very exhausting” to “very easy”.}~\cite{Liao2018TheNT, Sawada2016TheNT}. 

The detailed MOS test scores are shown in Fig.~\ref{fig:inpara_mos_4dims}. We observe that the MOS scores difference is not big, which was similarly observed in~\cite{Hu2016DiscoursePA}. We conjecture that may be due to the fact that long-form paragraph samples are too long for subjects to remember all the differences for a clearly distinguishable score. The results can still shed some light on the power of \textbf{ParaTTS}, which obtains scores consistently higher than the \textbf{Baseline} in all four testing fronts, where better naturalness and pleasantness are also reflected in the lower MCDs and higher LF0 correlations. The pause (break) MOS score of the \textbf{ParaTTS} is almost close to that of the recording, which is also confirmed with lower RMSE in pause duration. Listening to the multi-sentence paragraph audios synthesized by the \textbf{ParaTTS} does not increase the listening fatigue of the listeners and gets an on-par listening comfort score with the recordings, possibly helped by its naturalness and pleasantness close to the recordings.

	\begin{table}[h]
		\caption{The information of in-domain and out-of-domain test sets. } 
		\label{tab:ood_test}
		\centering
		\small
		\setlength{\tabcolsep}{3.5pt}
		\begin{tabular}{|p{{55pt}}<{\centering}|p{47pt}<{\centering}|p{44pt}<{\centering}|p{40pt}<{\centering}|p{28pt}<{\centering}|}
			\hline		
			\multirow{2}*{Domain}   & Paragraph     & \multirow{2}*{\#Paragraphs}        & \multirow{2}*{\#Sentences}      & \multirow{2}*{Minutes}             \\
			& length & & & \\ \hline
			In-domain               & Short                  & 32      & 5      & 0.5          \\ 	\cline{1-5}  
\multirow{2}*{Out-of-domain}        & Long                   & 12      & 23     & 2                          \\  \cline{2-5}
			                        & Extra-long             & 6       & 51     & 5          \\
			\hline
		\end{tabular}
	\end{table}

\begin{figure}[h]
    \begin{center}
     \includegraphics[scale=0.65]{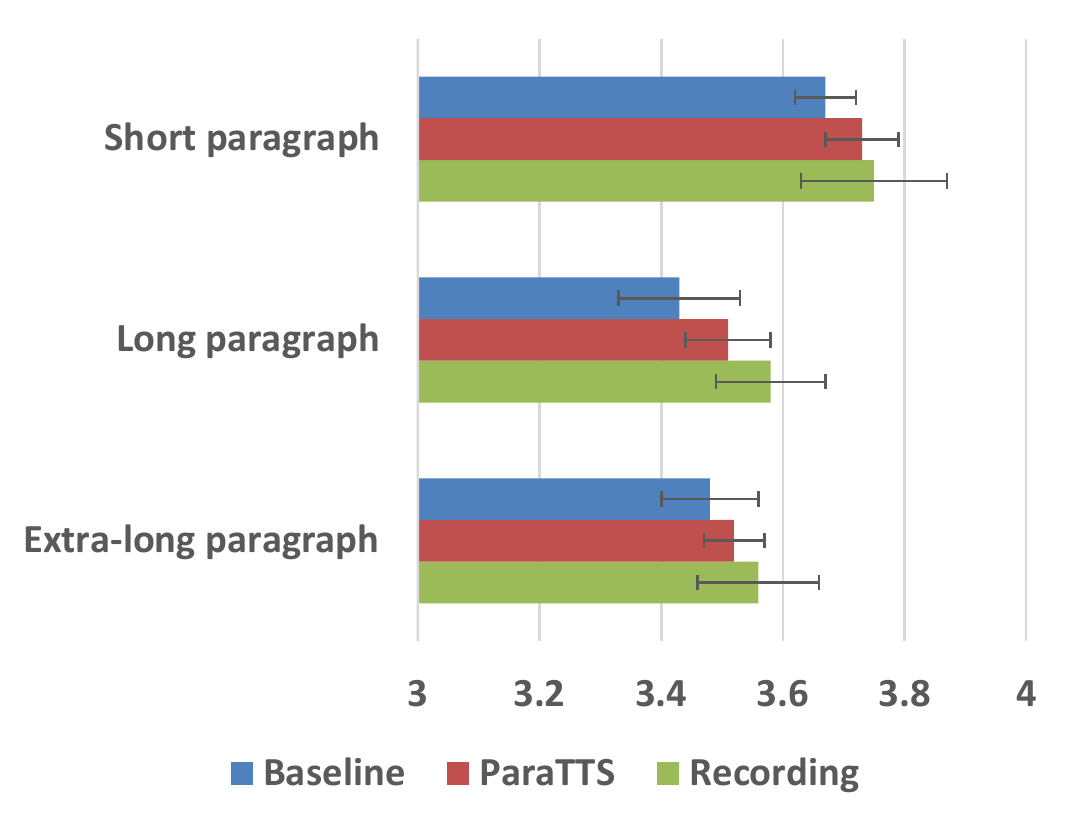}
    \end{center} \vspace*{-10pt}
    \caption{Overall impression MOS scores with 95\% confidence intervals on in-domain and out-domain test sets.}
    \label{fig:mos_overall_3domain}	 \vspace*{-10pt}
\end{figure}

\subsection{Out-of-domain Test}
To examine the proposed model's generalization ability, we extend the test to out-of-domain, long paragraphs and extra-long paragraphs. The information on the in-domain and out-of-domain test sets is listed in Table~\ref{tab:ood_test}. In addition to the in-domain 38 short paragraphs, there are 12 long paragraphs and 6 extra-long paragraphs. On the average, each paragraph contains 5, 23 and 51 utterances, respectively. The overall impression MOS results are shown in Fig.~\ref{fig:mos_overall_3domain}. It is observed that the MOS scores of the recordings decrease with the length of the paragraph increases. In other words, even for the original recorded speech, long paragraphs tend to be getting lower MOS scores or causing more listening fatigue than shorter paragraphs. In the extra-long paragraphs testing set, there is still occasional skipping issue even though we adopt the robust GMMv2b attention mechanism, indicating that it is still challenging for long-sequence modeling in the attention mechanism.

\textbf{ParaTTS} yields higher scores than the \textbf{Baseline}, not only for the in-domain short paragraphs but for the out-of-domain, long and extra-long paragraphs. In inference, the multi-head attention in the linguistics-aware network plays a role of a self-attention mechanism. The encoders in both the TTS backbone and linguistic-aware network take the paragraph phoneme sequence as input to exploit the contextual embedding vector for rendering more natural speech. In this way, the advantage of long-range dependency utilized by the self-attention mechanism generalizes the model for synthesizing longer paragraphs.	

\section{Conclusion} 
\label{sec:conclusion}
In this research, we propose to use a new, paragraph-based, end-to-end TTS model to model linguistic and prosodic information embedded in paragraph text with the corresponding acoustic data. We design both linguistics-aware and prosody-aware networks to learn the information via a paragraph encoder and its multi-head attention mechanism. Additionally, a sentence-position network is used to exploit the inter-sentence information in the paragraph. Trained on a storytelling, audio-book corpus (4.08 hours), recorded by a female Mandarin speaker, experimental results show that the proposed new paragraph-based model can produce TTS speech better than the conventional sentence-based TTS baseline system, both objectively and subjectively. The new model can learn the cross-sentence information well, e.g., the break durations between adjacent sentences, and generalize the learned information to longer or much longer paragraphs than those used in the training corpus.

	\bibliographystyle{IEEEtran}
	\bibliography{main}

\begin{thebibliography}{10}
\providecommand{\url}[1]{#1}
\csname url@samestyle\endcsname
\providecommand{\newblock}{\relax}
\providecommand{\bibinfo}[2]{#2}
\providecommand{\BIBentrySTDinterwordspacing}{\spaceskip=0pt\relax}
\providecommand{\BIBentryALTinterwordstretchfactor}{4}
\providecommand{\BIBentryALTinterwordspacing}{\spaceskip=\fontdimen2\font plus
\BIBentryALTinterwordstretchfactor\fontdimen3\font minus
  \fontdimen4\font\relax}
\providecommand{\BIBforeignlanguage}[2]{{%
\expandafter\ifx\csname l@#1\endcsname\relax
\typeout{** WARNING: IEEEtran.bst: No hyphenation pattern has been}%
\typeout{** loaded for the language `#1'. Using the pattern for}%
\typeout{** the default language instead.}%
\else
\language=\csname l@#1\endcsname
\fi
#2}}
\providecommand{\BIBdecl}{\relax}
\BIBdecl

\bibitem{Wang2017TacotronTE}
Y.~Wang, R.~Skerry-Ryan, D.~Stanton, Y.~Wu, R.~J. Weiss, N.~Jaitly, Z.~Yang,
  Y.~Xiao, Z.~Chen, S.~Bengio \emph{et~al.}, ``Tacotron: Towards end-to-end
  speech synthesis,'' \emph{Proc. Interspeech 2017}, pp. 4006--4010, 2017.

\bibitem{shen2018natural}
J.~Shen, R.~Pang, R.~J. Weiss, M.~Schuster, N.~Jaitly, Z.~Yang, Z.~Chen,
  Y.~Zhang, Y.~Wang, R.~Skerry-Ryan, R.~Saurous, Y.~Agiomyrgiannakis, and
  Y.~Wu, ``Natural {TTS} synthesis by conditioning wavenet on mel spectrogram
  predictions,'' in \emph{ICASSP}, 2018.

\bibitem{Li2019NeuralSS}
N.~Li, S.~Liu, Y.~Liu, S.~Zhao, and M.~Liu, ``Neural speech synthesis with
  transformer network,'' in \emph{AAAI}, 2019.

\bibitem{Ren2019FastSpeechFR}
Y.~Ren, Y.~Ruan, X.~Tan, T.~Qin, S.~Zhao, Z.~Zhao, and T.-Y. Liu, ``Fastspeech:
  fast, robust and controllable text to speech,'' in \emph{Proceedings of the
  33rd International Conference on Neural Information Processing Systems},
  2019, pp. 3171--3180.

\bibitem{Ren2020FastSpeech2F}
Y.~Ren, C.~Hu, X.~Tan, T.~Qin, S.~Zhao, Z.~Zhao, and T.-Y. Liu, ``Fastspeech 2:
  Fast and high-quality end-to-end text to speech,'' in \emph{ICLR}, 2020.

\bibitem{Yu2020DurIANDI}
C.~Yu, H.~Lu, N.~Hu, M.~Yu, C.~Weng, K.~Xu, P.~Liu, D.~Tuo, S.~Kang, G.~Lei,
  D.~Su, and D.~Yu, ``{DurIAN}: Duration informed attention network for speech
  synthesis,'' in \emph{INTERSPEECH}, 2020.

\bibitem{Oord2016WaveNetAG}
A.~van~den Oord, S.~Dieleman, H.~Zen, K.~Simonyan, O.~Vinyals, A.~Graves,
  N.~Kalchbrenner, A.~Senior, and K.~Kavukcuoglu, ``Wavenet: A generative model
  for raw audio,'' in \emph{9th ISCA Speech Synthesis Workshop}, pp. 125--125.

\bibitem{Valin2019LPCNETIN}
J.~Valin and J.~Skoglund, ``{LPCNET}: Improving neural speech synthesis through
  linear prediction,'' in \emph{ICASSP}, 2019.

\bibitem{Kumar2019MelGANGA}
K.~Kumar, R.~Kumar, T.~de~Boissiere, L.~Gestin, W.~Z. Teoh, J.~Sotelo,
  A.~de~Brebisson, Y.~Bengio, and A.~Courville, ``{MelGAN}: generative
  adversarial networks for conditional waveform synthesis,'' in
  \emph{Proceedings of the 33rd International Conference on Neural Information
  Processing Systems}, 2019, pp. 14\,910--14\,921.

\bibitem{Kong2020HiFiGANGA20hifi}
J.~Kong, J.~Kim, and J.~Bae, ``{HiFi}-{GAN}: Generative adversarial networks
  for efficient and high fidelity speech synthesis,'' \emph{NIPS}, 2020.

\bibitem{granville1993algorithm}
R.~Granville, ``An algorithm for high-level organization of multi-paragraph
  texts,'' in \emph{Intentionality and Structure in Discourse Relations}, 1993.

\bibitem{hearst1994multi}
M.~A. Hearst, ``Multi-paragraph segmentation of expository text,'' in
  \emph{Proceedings of the 32nd annual meeting on Association for Computational
  Linguistics}, 1994, pp. 9--16.

\bibitem{Smith2004TopicTA}
C.~Smith, ``Topic transitions and durational prosody in reading aloud:
  production and modeling,'' \emph{Speech Communication}, vol.~42, pp.
  247--270, 2004.

\bibitem{Tseng2006ProsodicFA}
C.-y. Tseng, Z.-y. Su, C.-H. Chang, and C.-h. Tai, ``Prosodic fillers and
  discourse markers--discourse prosody and text prediction,'' in \emph{Tonal
  Aspects of Languages}, 2006.

\bibitem{Cole2015ProsodyIC}
J.~Cole, ``Prosody in context: a review,'' \emph{Language, Cognition and
  Neuroscience}, vol.~30, pp. 1 -- 31, 2015.

\bibitem{Kreiman1982PerceptionOS}
J.~Kreiman, ``Perception of sentence and paragraph boundaries in natural
  conversation,'' \emph{Journal of Phonetics}, vol.~10, no.~2, pp. 163--175,
  1982.

\bibitem{Grosz1992SomeIC}
B.~Grosz and J.~Hirschberg, ``Some intonational characteristics of discourse
  structure,'' in \emph{ICSLP}, 1992.

\bibitem{Hirschberg1996APA}
J.~Hirschberg and C.~H. Nakatani, ``A prosodic analysis of discourse segments
  in direction-giving monologues,'' in \emph{ACL}, 1996.

\bibitem{Swerts1994ProsodyAA}
M.~Swerts and R.~Geluykens, ``Prosody as a marker of information flow in spoken
  discourse,'' \emph{Language and Speech}, vol.~37, pp. 21 -- 43, 1994.

\bibitem{alias2005high}
F.~Al{\i}as, I.~Iriondo, L.~Formiga, X.~Gonzalvo, C.~Monzo, and X.~Sevillano,
  ``High quality spanish restricted-domain {TTS} oriented to a weather forecast
  application,'' in \emph{INTERSPEECH}, 2005.

\bibitem{sarkar2015modeling}
P.~Sarkar and K.~S. Rao, ``Modeling pauses for synthesis of storytelling style
  speech using unsupervised word features,'' \emph{Procedia Computer Science},
  vol.~58, pp. 42--49, 2015.

\bibitem{sarkar2015data}
{Sarkar, Parakrant and Rao, K Sreenivasa}, ``Data-driven pause prediction for
  synthesis of storytelling style speech based on discourse modes,'' in
  \emph{CONECCT}, 2015.

\bibitem{sarkar2015analysis}
{{Sarkar, Parakrant and Rao, K Sreenivasa}}, ``Analysis and modeling pauses for
  synthesis of storytelling speech based on discourse modes,'' \emph{IC3},
  2015.

\bibitem{tyagi2020dynamic}
S.~Tyagi, M.~Nicolis, J.~Rohnke, T.~Drugman, and J.~Lorenzo-Trueba, ``Dynamic
  prosody generation for speech synthesis using linguistics-driven acoustic
  embedding selection,'' \emph{Proc. Interspeech 2020}, pp. 4407--4411, 2020.

\bibitem{zhang2004prominence}
J.~Y. Zhang, A.~R. Toth, K.~Collins-Thompson, and A.~W. Black, ``Prominence
  prediction for supersentential prosodic modeling based on a new database,''
  in \emph{SSW}, 2004.

\bibitem{prahallad2006sub}
K.~Prahallad, A.~W. Black, and R.~Mosur, ``Sub-phonetic modeling for capturing
  pronunciation variations for conversational speech synthesis,'' in \emph{2006
  IEEE International Conference on Acoustics Speech and Signal Processing
  Proceedings}, vol.~1.\hskip 1em plus 0.5em minus 0.4em\relax IEEE, 2006, pp.
  I--I.

\bibitem{bennett2005prediction}
C.~L. Bennett and A.~W. Black, ``Prediction of pronunciation variations for
  speech synthesis: A data-driven approach,'' in \emph{ICASSP}, 2005.

\bibitem{miller1998pronunciation}
C.~A. Miller, \emph{Pronunciation modeling in speech synthesis}.\hskip 1em plus
  0.5em minus 0.4em\relax University of Pennsylvania, 1998.

\bibitem{Battenberg2020LocationRelativeAM}
E.~Battenberg, R.~Skerry-Ryan, S.~Mariooryad, D.~Stanton, D.~Kao, M.~Shannon,
  and T.~Bagby, ``Location-relative attention mechanisms for robust long-form
  speech synthesis,'' in \emph{ICASSP}, 2020.

\bibitem{Wang2020sTransformerSF}
X.~Wang, H.~Ming, L.~He, and F.~Soong, ``s-transformer: Segment-transformer for
  robust neural speech synthesis,'' \emph{ArXiv}, vol. abs/2011.08480, 2020.

\bibitem{Hu2016DiscoursePA}
N.~Hu, P.~Shao, Y.~Zu, Z.~Wang, W.~Huang, and S.~Wang, ``Discourse prosody and
  its application to speech synthesis,'' in \emph{ISCSLP}, 2016.

\bibitem{Peir-Lilja2018}
A.~{Peiró-Lilja} and M.~Farrús, ``Paragraph prosodic patterns to enhance
  text-to-speech naturalness,'' in \emph{Speech Prosody}, 2018, pp. 612--616.

\bibitem{Clark2019EvaluatingLT}
R.~Clark, H.~Silen, T.~Kenter, and R.~Leith, ``Evaluating long-form
  text-to-speech: Comparing the ratings of sentences and paragraphs,'' in
  \emph{Proc. 10th ISCA Speech Synthesis Workshop}, pp. 99--104.

\bibitem{Aubin2019ImprovingSS}
A.~Aubin, A.~Cervone, O.~Watts, and S.~King, ``Improving speech synthesis with
  discourse relations,'' in \emph{INTERSPEECH}, 2019.

\bibitem{Farrs2016ParagraphbasedPC}
M.~Farr{\'u}s, C.~Lai, and J.~Moore, ``Paragraph-based prosodic cues for speech
  synthesis applications,'' \emph{Speech prosody}, pp. 1143--1147, 2016.

\bibitem{pan2021chapter}
J.~Pan, L.~Wu, X.~Yin, P.~Wu, C.~Xu, and Z.~Ma, ``A chapter-wise understanding
  system for text-to-speech in chinese novels,'' in \emph{ICASSP}, 2021.

\bibitem{hodari2021camp}
Z.~Hodari, A.~Moinet, S.~Karlapati, J.~Lorenzo-Trueba, T.~Merritt, A.~Joly,
  A.~Abbas, P.~Karanasou, and T.~Drugman, ``{CAMP}: a two-stage approach to
  modelling prosody in context,'' in \emph{ICASSP}, 2021.

\bibitem{karlapati2021prosodic}
S.~Karlapati, A.~Abbas, Z.~Hodari, A.~Moinet, A.~Joly, P.~Karanasou, and
  T.~Drugman, ``Prosodic representation learning and contextual sampling tor
  neural text-to-speech,'' in \emph{ICASSP}, 2021.

\bibitem{xu2021improving}
G.~Xu, W.~Song, Z.~Zhang, C.~Zhang, X.~He, and B.~Zhou, ``Improving prosody
  modelling with cross-utterance bert embeddings for end-to-end speech
  synthesis,'' in \emph{ICASSP}, 2021.

\bibitem{guo2021conversational}
H.~Guo, S.~Zhang, F.~K. Soong, L.~He, and L.~Xie, ``Conversational end-to-end
  {TTS} for voice agents,'' in \emph{SLT)}, 2021.

\bibitem{kentonbert}
J.~D. M.-W.~C. Kenton and L.~K. Toutanova, ``{BERT}: Pre-training of deep
  bidirectional transformers for language understanding,'' in \emph{Proceedings
  of NAACL-HLT}, 2019, pp. 4171--4186.

\bibitem{Guo2019}
H.~Guo, F.~K. Soong, L.~He, and L.~Xie, ``Exploiting syntactic features in a
  parsed tree to improve end-to-end {TTS},'' \emph{Proc. Interspeech 2019}, pp.
  4460--4464, 2019.

\bibitem{oplustil2020using}
P.~Oplustil-Gallegos and S.~King, ``Using previous acoustic context to improve
  text-to-speech synthesis,'' \emph{arXiv preprint arXiv:2012.03763}, 2020.

\bibitem{oplustil2021comparing}
P.~Oplustil-Gallegos, J.~O’Mahony, and S.~King, ``Comparing acoustic and
  textual representations of previous linguistic context for improving
  text-to-speech,'' in \emph{SSW}, 2021.

\bibitem{lee2017fully}
J.~Lee, K.~Cho, and T.~Hofmann, ``Fully character-level neural machine
  translation without explicit segmentation,'' \emph{Transactions of the
  Association for Computational Linguistics}, vol.~5, pp. 365--378, 2017.

\bibitem{srivastava2015highway}
R.~K. Srivastava, K.~Greff, and J.~Schmidhuber, ``Highway networks,''
  \emph{arXiv preprint arXiv:1505.00387}, 2015.

\bibitem{chung2014empirical}
J.~Chung, C.~Gulcehre, K.~Cho, and Y.~Bengio, ``Empirical evaluation of gated
  recurrent neural networks on sequence modeling,'' in \emph{NIPS 2014 Workshop
  on Deep Learning, December 2014}, 2014.

\bibitem{bahdanau2014neural}
D.~Bahdanau, K.~H. Cho, and Y.~Bengio, ``Neural machine translation by jointly
  learning to align and translate,'' in \emph{3rd International Conference on
  Learning Representations, ICLR 2015}, 2015.

\bibitem{Vaswani2017AttentionIA}
A.~Vaswani, N.~Shazeer, N.~Parmar, J.~Uszkoreit, L.~Jones, A.~N. Gomez,
  {\L}.~Kaiser, and I.~Polosukhin, ``Attention is all you need,'' in
  \emph{Proceedings of the 31st International Conference on Neural Information
  Processing Systems}, 2017, pp. 6000--6010.

\bibitem{SkerryRyan2018TowardsEP}
R.~Skerry-Ryan, E.~Battenberg, Y.~Xiao, Y.~Wang, D.~Stanton, J.~Shor, R.~Weiss,
  R.~Clark, and R.~A. Saurous, ``Towards end-to-end prosody transfer for
  expressive speech synthesis with {Tacotron},'' in \emph{ICML}, 2018.

\bibitem{young2002htk}
S.~J. Young, J.~Jansen, J.~J. Odell, D.~Ollason, and P.~C. Woodland, ``The
  {HTK} book,'' 1995.

\bibitem{kingma2014adam}
D.~P. Kingma and J.~Ba, ``Adam: A method for stochastic optimization,'' in
  \emph{ICLR}, 2015.

\bibitem{benesty2009pearson}
J.~Benesty, J.~Chen, Y.~Huang, and I.~Cohen, ``Pearson correlation
  coefficient,'' in \emph{Noise reduction in speech processing}.\hskip 1em plus
  0.5em minus 0.4em\relax Springer, 2009, pp. 1--4.

\bibitem{Ouden2009ProsodicRO}
H.~Den~Ouden, L.~Noordman, and J.~Terken, ``Prosodic realizations of global and
  local structure and rhetorical relations in read aloud news reports,''
  \emph{Speech Communication}, vol.~51, no.~2, pp. 116--129, 2009.

\bibitem{Theune2006GeneratingES}
M.~Theune, K.~Meijs, D.~Heylen, and R.~Ordelman, ``Generating expressive speech
  for storytelling applications,'' \emph{IEEE Transactions on Audio, Speech,
  and Language Processing}, vol.~14, pp. 1137--1144, 2006.

\bibitem{Liao2018TheNT}
Y.-F. Liao, Y.-B. Chai, and C.-H. Tsai, ``The {NTUT}’s text-to-speech system
  for blizzard challenge 2018,'' 2018.

\bibitem{Sawada2016TheNT}
K.~Sawada, C.~Asai, K.~Hashimoto, K.~Oura, and K.~Tokuda, ``The {NITech}
  text-to-speech system for the blizzard challenge 2016,'' 2016.

\end{thebibliography}

\end{document}